\documentclass[fleqn,usenatbib]{mnras}
\usepackage{newtxtext,newtxmath}
\usepackage{amsmath}
\usepackage{graphics,graphicx}
\usepackage{natbib}
\usepackage{setspace}
\usepackage{caption,setspace}

\usepackage{subcaption}
\usepackage{multicol}

\usepackage{mathtools}
\usepackage{enumitem}
\usepackage{lipsum}
\setlist[itemize]{leftmargin=*}

\usepackage{floatrow}
\usepackage[outercaption]{sidecap}    
\sidecaptionvpos{figure}{t}

\usepackage[T1]{fontenc}
\usepackage{ae,aecompl}
\pdfoutput=1

\title[Predicting sub-millimeter flux densities from global galaxy properties]{Predicting sub-millimeter flux densities from global galaxy properties}
\author[R.K. Cochrane et al.]{R. K. Cochrane, $^{1,2}$\thanks{E-mail: rcochrane@flatironinstitute.org}
C. C. Hayward$^{1}$, D. Angl\'es-Alc\'azar$^{1,3}$
\& R. S. Somerville$^{1}$
\\
$^{1}$Center for Computational Astrophysics, Flatiron Institute, 162 Fifth Avenue, New York, NY 10010, USA\\
$^{2}$Harvard-Smithsonian Center for Astrophysics, 60 Garden St. Cambridge, MA 02138, USA\\
$^{3}$Department of Physics, University of Connecticut, 196 Auditorium Road, U-3046, Storrs, CT 06269-3046, USA}
\date{Accepted XXX. Received YYY; in original form ZZZ}

\pubyear{2022}
\begin{document}
\label{firstpage}
\pagerange{\pageref{firstpage}--\pageref{lastpage}}
\maketitle
\begin{abstract}
Recent years have seen growing interest in post-processing cosmological simulations with radiative transfer codes to predict observable fluxes for simulated galaxies. However, this can be slow, and requires a number of assumptions in cases where simulations do not resolve the ISM. Zoom-in simulations better resolve the detailed structure of the ISM and the geometry of stars and gas, however statistics are limited due to the computational cost of simulating even a single halo. In this paper, we make use of a set of high resolution, cosmological zoom-in simulations of massive ($M_{\star}\gtrsim10^{10.5}\,\rm{M_{\odot}}$ at $z=2$), star-forming galaxies from the FIRE suite. We run the {\sc skirt} radiative transfer code on hundreds of snapshots in the redshift range $1.5<z<5$ and calibrate a power law scaling relation between dust mass, star formation rate and $870\,\mu\rm{m}$ flux density. The derived scaling relation shows encouraging consistency with observational results from the sub-millimeter-selected AS2UDS sample. We extend this to other wavelengths, deriving scaling relations between dust mass, stellar mass, star formation rate and redshift and sub-millimeter flux density at observed-frame wavelengths between $\sim340\,\mu\rm{m}$ and $\sim870\,\mu\rm{m}$. We then apply the scaling relations to galaxies drawn from EAGLE, a large box cosmological simulation. We show that the scaling relations predict EAGLE sub-millimeter number counts that agree well with previous results that were derived using far more computationally expensive radiative transfer techniques. Our scaling relations can be applied to other simulations and semi-analytical or semi-empirical models to generate robust and fast predictions for sub-millimeter number counts. 
\end{abstract}
\section{Introduction}
Robust constraints on the history of star formation in the Universe are fundamental to studies of galaxy evolution. Rest-frame ultraviolet (UV) surveys, particularly in the Hubble Deep Fields, have measured unobscured star formation out to $z\sim9$ and beyond \citep[e.g.][]{Oesch2018,Bouwens2019,Bouwens2021a,Bouwens2022,Stefanon2021a,Finkelstein2022,Tacchella2022}. However, such surveys are biased towards the least dust-obscured systems, which may represent a sub-dominant population at certain cosmic epochs. The population of rare but bright sub-millimeter galaxies (SMGs, \citealt{Smail1997,Barger1998,Hughes1998,Blain2002}) discovered by early single-dish sub-millimeter observations appears to contribute significantly to the star formation rate density of the Universe (SFRD). \cite{Zavala2021} estimate that dust-obscured star formation dominates the SFRD out to $z\sim4$, and roughly equals the contribution of unobscured star formation (traced by the rest-frame UV and optical emission) at $z\sim4-5$ \citep[see also][]{Swinbank2014,Dunlop2017,Casey2018c,Bouwens2020a}. \\
\indent The most extreme SMGs ($L_{\rm{IR}} \sim 10^{13}\,\rm{L}_{\odot}$) form stars at rates of thousands of solar masses per year, have established high stellar masses ($M_{\star}\sim10^{11}\,\rm{M_{\odot}}$; \citealt{Magnelli2012,Simpson2014,Cunha2015,Miettinen2017,Dudzeviciute2019}) within the first few Gyr of cosmic time, and reside in the most massive haloes \citep{Hickox2012,Wilkinson2016,Marrone2018,Miller2018,Garcia-Vergara2020,Stach2021}. They hence place important constraints on the formation of dust and metals in the early Universe \citep{Rowlands2014,Magnelli2020}. However, detailed study of these systems has historically been difficult, due in part to the poor spatial resolution of single dish telescopes such as SCUBA and the lack of deep multi-wavelength data for cross-matching sub-millimeter detections (often of blended sources) to counterparts at other wavelengths. \\
\indent However, recent developments in sub-millimeter interferometry, most recently the depth and angular resolution brought by ALMA, has transformed our understanding of SMGs \citep[see the comprehensive review by][]{Hodge2020}. We can now locate individual sources precisely, and derive much stronger constraints on source multiplicity (e.g. \citealt{Stach2018} find that $28\pm2\%$ of single dish sources with $S_{850}\geq 5\,\rm{mJy}$ are blends of two or more ALMA-detectable SMGs, each with flux density $\geq 1\,\rm{mJy}$; see also \citealt{Hayward2013,Hayward2013b,Hayward2018a}). This has led to more robust SMG number counts, redshift distributions, and counterpart identification, which has enabled estimates of physical properties \citep[e.g.][]{Dudzeviciute2019}. We can now resolve the dust continuum and line emission of distant galaxies on sub-kiloparsec scales (and even lower for some strongly-lensed sources; e.g. \citealt{Rybak2015a,Rybak2015,Rybak2020,Fujimoto2021}). A number of studies have reported that observed dust continuum emission is not co-located with emission in the rest-frame UV and optical \citep[e.g.][]{Chen2017,Cochrane2021}, with the longer-wavelength emission typically being more compact (e.g. $1-2\,\rm{kpc}$;  \citealt{Hodge2016,Tadaki2017,Rivera2018,Chen2020}). In some high-redshift sources, dust attenuation is high enough to render the sources near-infrared-faint \citep[see][for a statistical study]{Smail2020}.\\
\indent Despite these major observational strides, SMGs present a challenge to models of galaxy formation. The majority of semi-analytic models (SAMs) fail to match the number counts of SMGs, under-predicting their numbers by 1-2 orders of magnitude \citep[e.g.][]{Cole2000,Granato2000,Somerville2012}. Previously explored solutions have involved invoking different variants of a top-heavy IMF \citep{Baugh2005,Lacey2016,Cowley2018}; adopting a particularly extreme flat IMF in starbursts enabled \cite{Baugh2005} to match observed SMG number counts and redshift distribution as well as the local $K-$band luminosity function. However, given the wide range of tuneable parameters in SAMs, it remains unclear whether this is the solution \citep{Safarzadeh2017}. One fundamental limitation is the lack of geometrical information in SAMs; while galaxies can be parametrised by, for example, a disk plus bulge component on which post-processing may be applied, the detailed relative geometries of dust and stars are not captured. \\
\indent Hydrodynamical simulations track the positions of stellar and gas (and sometimes dust) particles, and hence do contain some information about their relative spatial distributions. There has been substantial recent interest in casting these simulations into the `observational plane', made possible by radiative transfer codes such as {\sc skirt} \citep{Baes2011,Camps2014}, {\sc powderday} \citep{Narayanan2020} and {\sc sunrise} \citep{Jonsson2010}. These codes work by propagating photons from ionising sources through a dust grid. Following substantial success in applying these codes to individual galaxies and limited samples of zoom-in simulations \citep[e.g.][]{Narayanan2010,Narayanan2015,Hayward2011,Hayward2012,Cochrane2019,Cochrane2022a,Parsotan2021}, recent work has focused on application to large scale cosmological simulations. The EAGLE project presents predictions for galaxy spectral energy distributions (SEDs) using {\sc skirt} \citep{Camps2016,Trayford2017}. \cite{McAlpine2019} show the good agreement between the physical properties (including the redshift distribution, SFRs, stellar and dust masses) of modelled and observed sub-mm sources (defined as those with $S_{850}\geq1\rm{mJy}$). However, the number counts of the most extreme sources are at least an order of magnitude lower than inferred from observations \citep{Cowley2018,Wang2019c}. Radiative transfer calculations (with {\sc powderday}) have also been performed on the S{\sc imba} simulations \citep{Dave2019}. \cite{Lovell2020} show that their model is more successful at matching observed $850\,\mu\rm{m}$ number counts than EAGLE (though the redshift distribution of bright $>3.6\,\rm{mJy}$ sources is skewed to higher redshifts than is currently measured by observations). S{\sc imba}'s larger number of sub-millimeter bright sources is tied to the `on-the-fly' dust model, which produces galaxies with dust masses that are higher by a factor of $\sim2.5$ than those calculated using a more commonly-adopted dust-to-metals ratio of $0.3\,\rm{dex}$. It is also related to the distribution of star formation rates at different redshifts: S{\sc imba} models more high mass, high specific star formation rate galaxies than EAGLE at $z\sim2$ and by $z>4$ the global star formation rate density is higher than inferred from observations \citep{Dave2019}. These existing works highlight the utility of SMG number counts as one of numerous observational constraints for the various physical processes included in galaxy formation models, such as AGN and stellar feedback and dust physics.\\ 
\indent One key limitation of applying radiative transfer to cosmological simulations is the computational resources required. For a project such as CAMELS \citep{Villaescusa-Navarro2021,Villaescusa-Navarro2022b}, which involves thousands of simulations with different initial conditions and parameter choices, this is intractable. Furthermore, for large-box simulations like EAGLE, the resolution is low and particle mass is high ($\sim10^{6}\,\rm{M_{\odot}}$), such that the cold molecular gas phase of the ISM is not resolved, and young stars can appear `clumped' into a few, point-like regions \citep{Camps2016}. It is therefore necessary to artificially increase the particle mass and time resolution by performing a resampling procedure on the gas and young star particles prior to the radiative transfer procedure. \\
\indent An alternative to running radiative transfer on large cosmological simulations is to apply well-calibrated scaling relations to the physical properties of simulated galaxies to estimate their fluxes. \cite{Hayward2011} used hydrodynamic simulations of isolated and merging galaxies to derive a relation between dust mass, SFR and $S_{850}$ \citep[see also][]{Hayward2013}. While a parametrised relation would be harder to derive for emission at shorter wavelengths, which have more complex dependencies on stellar populations and geometry/inclination, this works fairly well for sub-mm fluxes, which are expected to depend fundamentally on dust mass and dust temperature. Such a relation can then be applied to cosmological simulations (with lower particle resolution) to predict sub-mm flux densities with a trivial computational cost. \cite{Hayward2021} applied the scaling relation derived in \cite{Hayward2013} to derive SMG number counts for galaxies in the original Illustris \citep{Genel2014,Vogelsberger2014a} and more recent IllustrisTNG simulations \citep{Pillepich2018,Weinberger2017}. Their predicted number counts for the two simulations differed significantly, with IllustrisTNG hosting fewer SMGs than Illustris by an order of magnitude. They argued that this was due to the lower dust masses and SFRs of high mass galaxies in IllustrisTNG. These differences in demographics relate to the differing subgrid recipes in the two simulations, and may relate to early, efficient galaxy quenching in IllustrisTNG. This example again highlights the sensitivity (and constraining power) of SMG number counts and redshift distributions to the subgrid models employed in cosmological simulations.\\
\indent The Feedback In Realistic Environments project \citep{Hopkins2014,Hopkins2017} provides an ideal set of high resolution zoom-in simulations with which to derive new scaling relations. Crucially, these simulations resolve the formation of giant molecular clouds, which is currently not possible over cosmological volumes. The implementation of radiative transfer does not, therefore, require particle resampling, and the detailed geometry of stars and dust can be better resolved (assuming that the dust traces the gas with a fixed dust-to-metals mass ratio). Coupling with radiative transfer models has shown that these simulations produce galaxies with realistic sizes in the sub-mm \citep{Cochrane2019}. The focus of this paper is on deriving updated scaling relations to predict sub-mm flux densities using the FIRE-2 simulations.\\
\indent The structure of this paper is as follows. In Section \ref{sims_intro}, we describe the FIRE simulations in detail and outline the radiative transfer methods used to make predictions for the rest-frame FIR emission of each simulation snapshot. In Section \ref{sec:scaling_relations}, we show that the predicted scaling relations between dust mass, star formation rate and observable sub-mm flux density are broadly consistent with a large, homogeneously-selected observational sample of SMGs. We then use our simulations to derive a simple, analytical formula that enables us to make rapid predictions for sub-mm flux density from SFR and dust mass. We validate this new relation using the observational data. In Section \ref{sec:four_flux_densities}, we extend this study to predict sub-mm flux densities at various wavelengths by also including stellar mass and redshift terms. We apply the model to galaxies in the EAGLE simulations in Section \ref{sec:eagle_application} and draw conclusions in Section \ref{sec:conclusions}.

\begin{figure*}
    \centering
	\includegraphics[scale=0.51]{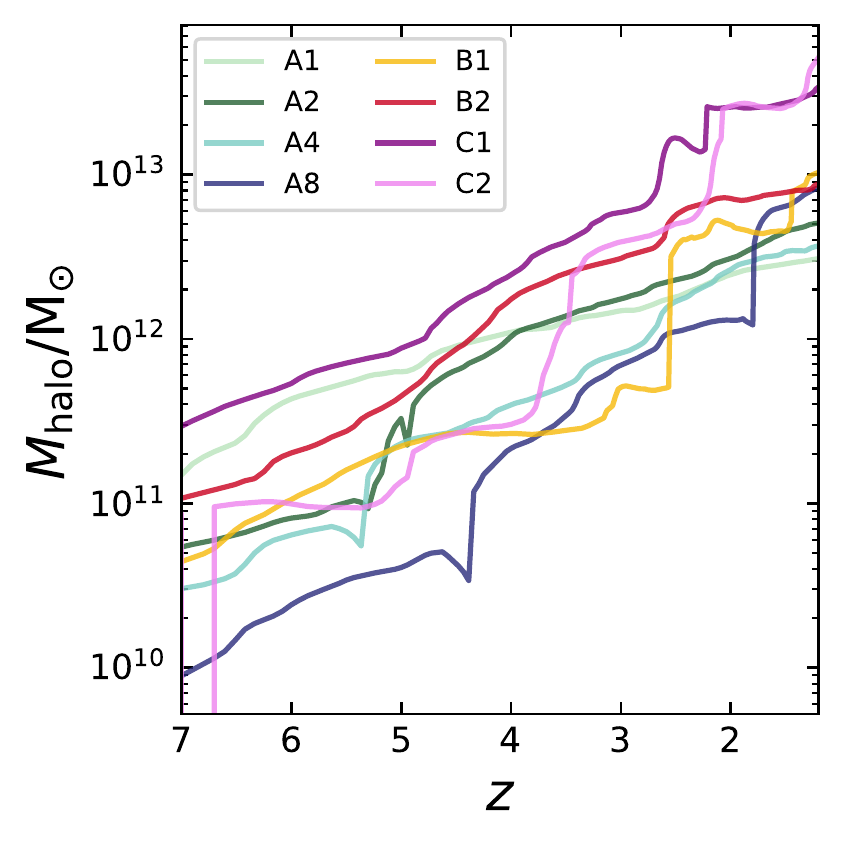}
	\includegraphics[scale=0.51]{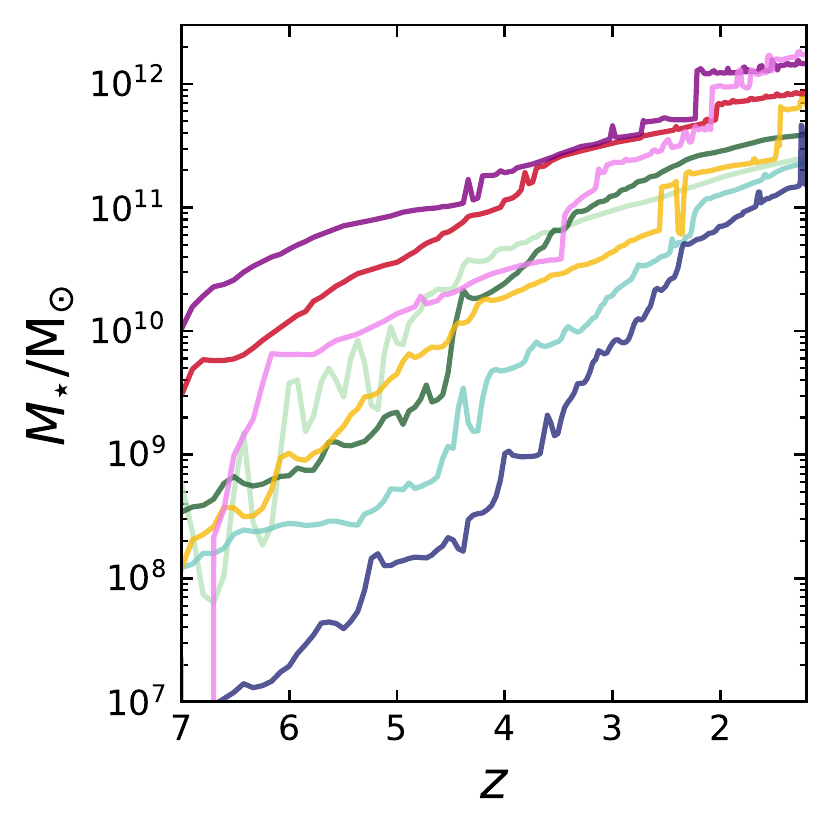}
	\includegraphics[scale=0.51]{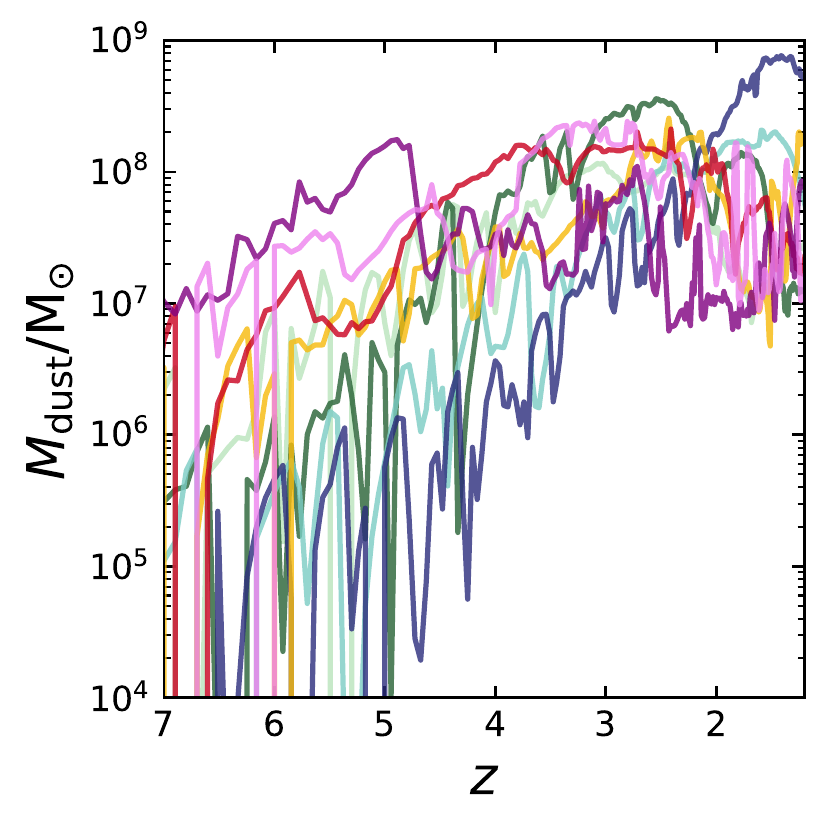}
	\includegraphics[scale=0.51]{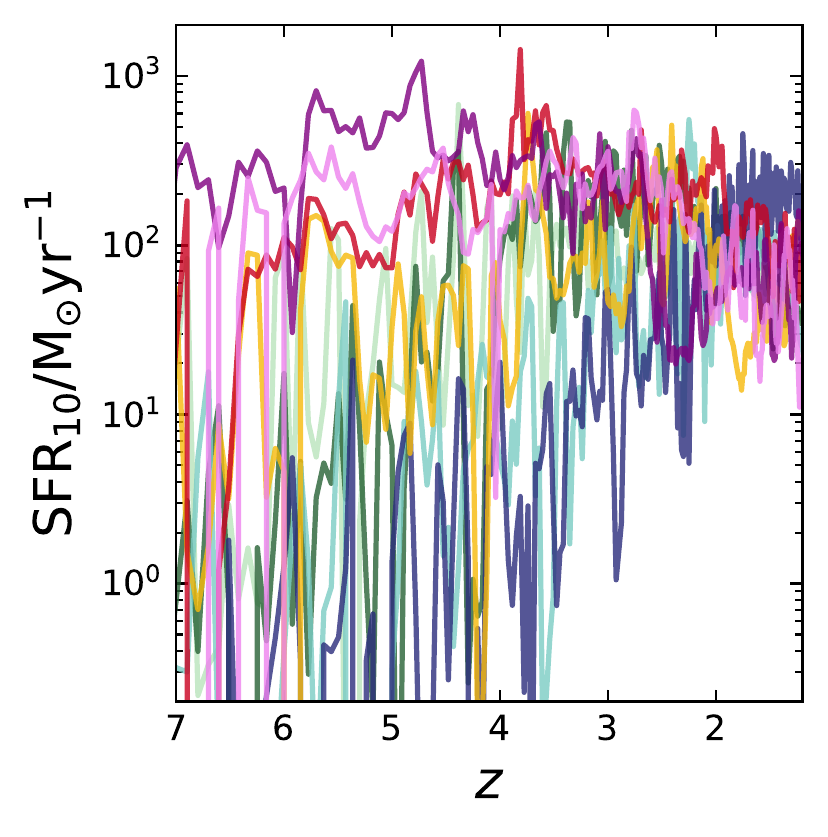}
	\caption{The evolution of physical quantities of the eight FIRE-2 haloes used in this paper. From left to right: evolution of halo mass; evolution of stellar mass within $0.1\,R_{\rm{vir}}$; evolution of dust mass within $0.1\,R_{\rm{vir}}$ (note that this is calculated by weighting the mass of metals in the gas phase by a constant dust-to-metals mass ratio); evolution of $10\,\rm{Myr}$-averaged star formation rate. The haloes display a variety of dark matter and stellar mass growth histories. Haloes C1 and C2 are the most massive, reaching $M_{\rm{halo}}>10^{13}\,\rm{M}_{\odot}$ by $z=2$.}
    \label{fig:evolution_halos}
\end{figure*}

\begin{table*}
\begin{center}
\begin{tabular}{c|c|c|c|c|c|c}
Name & $\log_{10}M_{\rm{halo}}/\rm{M_{\odot}}$ & $\log_{10}M_{\rm{\star}}/\rm{M_{\odot}}$ & $\log_{10}M_{\rm{gas}}/\rm{M_{\odot}}$ & $\log_{10}M_{\rm{dust}}/\rm{M_{\odot}}$ & $\rm{SFR_{10}}/\rm{M_{\odot}}\rm{yr}^{-1}$ & $S_{870}/\rm{mJy}$\\ 
\hline
A1 & $12.17$ & $10.98$ & $10.11$ & $8.04$ & $133$ & $0.65$ \\
A2 & $12.24$ & $11.09$ & $10.53$ & $8.41$ & $57$ & $0.95$  \\
A4 & $11.91$ & $10.29$ & $10.22$ & $7.70$ & $61$ & $0.23$  \\
A8 & $11.84$ & $9.90$ & $10.11$ & $7.50$ & $55$ & $0.16$  \\
B1 & $11.71$ & $10.64$ & $10.19$ & $7.78$ & $46$ & $0.32$  \\
B2 & $12.49$ & $11.52$ & $10.63$ & $8.16$ & $261$ & $1.19$  \\
C1 & $12.77$ & $11.66$ & $10.62$ & $7.76$ & $382$ & $0.80$  \\
C2 & $12.59$ & $11.36$ & $10.50$ & $8.33$ & $358$ & $1.60$  \\
\hline
\end{tabular}
\caption{Properties of the eight simulated FIRE-2 haloes at $z=3$. $M_{\rm{halo}}$ denotes the total mass of dark matter within the halo's virial radius, $R_{\rm{vir}}$. $M_{\rm{\star}}$, $M_{\rm{gas}}$, and $\rm{SFR_{10}}$ denote the stellar mass, gas mass, $10\,\rm{Myr}$ average star formation rate, and observed-frame $870\,\mu\rm{m}$ flux density of the halo's central galaxy, all calculated within $0.1R_{\rm{vir}}$.}
\label{Table:FIRE_halos}
\end{center}
\end{table*}

\section{Hydrodynamical simulations and radiative transfer calculations}\label{sims_intro}
\subsection{Galaxies in the FIRE-2 simulations}\label{sec:FIRE2}
The FIRE project \citep{Hopkins2014,Hopkins2017} is a suite of state-of-the-art hydrodynamical cosmological zoom-in simulations designed to explore the role of stellar feedback in galaxy formation and evolution. Without stellar feedback, the ISM would collapse on dynamical timescales, leading to accelerated star formation. This is ruled out by the gas consumption timescales of observed galaxies, as well as by galaxy stellar mass functions and the stellar mass-halo mass relation (see the overviews presented in \citealt{Hopkins2011} and \citealt{Somerville}). Galactic outflows are believed to regulate the mass of galaxies over time \citep{Keres2009,Faucher-Giguere2015,Angles-Alcazar2016} and to enrich the circumgalactic medium \citep[see][]{Hafen2019} and intergalactic medium (see the review by \citealt{Rupke2018}). Numerous stellar feedback processes are believed to contribute, in a complex, non-linear manner. These include supernovae, protostellar jets, photo-heating, stellar mass loss from O- and AGB-stars and radiation pressure \citep[see the review of][]{Dale2015}. Cosmological zoom simulations have only recently achieved sufficient resolution to model these feedback processes directly.\\
\indent Multi-channel feedback is modelled by FIRE using two primary techniques. Firstly, FIRE resolves the formation of giant molecular clouds (GMCs), with star formation taking place only in self-gravitating (according to the \citealt{Hopkins2013sf_criteria} criterion), self-shielding molecular gas (following \citealt{Krumholz2011}) at high densities ($n_{H}>1000\,\rm{cm}^{-3}$ in the simulations we use here). Secondly, FIRE includes models for energy and momentum return from the main stellar feedback processes, with direct implementation of the predictions of stellar population synthesis models without parameter tuning. Once a star particle forms, the simulations explicitly follow several different stellar feedback mechanisms, as detailed in \cite{Hopkins2018feedback}, including (1) local and long-range momentum flux from radiation pressure (in both the initial UV/optical single-scattering regime and re-radiated light in the IR); (2) energy, momentum, mass and metal injection from SNe (Types Ia and II) and stellar mass loss (both OB and AGB); and (3) photo-ionization and photo-electric heating. Every star particle is treated as a single stellar population with known mass, age, and metallicity. All feedback event rates, luminosities and energies, mass-loss rates, and all other quantities are tabulated directly from stellar evolution models ({\sc starburst99}; \citealt{Leitherer1999}), assuming a \citet{Kroupa2002} initial mass function. The FIRE prescription enables the self-consistent generation of galactic winds \citep{Muratov2015a,Angles-Alcazar2016,Pandya2021} and the reproduction of many observed galaxy properties, including stellar masses, star formation histories and the `main sequence' of star-forming galaxies (see \citealt{Hopkins2014,Sparre2017}), metallicities and abundance ratios \citep{Ma2016, VandeVoort2015}, as well as morphologies and kinematics of both thin and thick disks \citep{Ma2017}. \\
\indent For this paper, we study the central galaxies of eight massive haloes originally selected and simulated by \citet{Feldmann2016, Feldmann2017a} with the original FIRE model \citep{Hopkins2014} as part of the {\sc MassiveFIRE} suite. The first four of these are drawn from the `A-series' of \cite{Feldmann2017a} (A1, A2, A4 and A8). These haloes were selected to have dark matter halo masses of $M_{\rm{halo}}\sim10^{12.5}\,\rm{M_{\odot}}$ at $z=2$. The central galaxies of these haloes have stellar masses of $7\times10^{10}-3\times10^{11}\,\rm{M_{\odot}}$ at $z=2$ (detailed in Table \ref{Table:FIRE_halos}), with a variety of formation histories; see \cite{Feldmann2017a} for details. The subset of four A-series haloes studied in this paper are drawn from \cite{Angles-Alcazar2017}, who re-simulated them down to $z=1$ with the upgraded FIRE-2 physics model \citep{Hopkins2017} and with a novel on-the-fly treatment for the seeding and growth of supermassive black holes (SMBHs; see \cite{Angl2018} for details of the gravitational torque-driven model). Compared to FIRE-1, FIRE-2 simulations are run with a new, more accurate hydrodynamics solver (a mesh-free Godunov solver implemented in the {\sc gizmo}\footnote{\url{http://www.tapir.caltech.edu/~phopkins/Site/GIZMO.html}} code; \citealt{Gaburov2011,Hopkins2015}). They also feature improved treatments of cooling and recombination rates, gravitational softening and numerical feedback coupling, and they adopt a higher density threshold for star formation \citep{Hopkins2018feedback}. Note that these simulations do not include feedback from the accreting SMBHs \footnote{Given that in extreme cases, AGN can be the dominant driver of cold dust emission \citep{McKinney2021}, it would be interesting to repeat our analyses with simulations including AGN feedback and emission in future work. However, we expect this effect to be unimportant for the bulk of the population.}. A small number of snapshots from these haloes were also studied in \cite{Cochrane2019} and \cite{Parsotan2021}. For this study, we supplement the four `A-series' haloes by re-running four more haloes from \cite{Feldmann2017a}, with the updated FIRE-2 physics. Two of the haloes are drawn from their `B-series' (B1 and B2) and two from the `C-series' (Cm1:0, hereafter C1, and C2:0, hereafter C2). The B and C series haloes were selected to have dark matter halo masses of $M_{\rm{halo}}\sim10^{13}\,\rm{M_{\odot}}$ and $M_{\rm{halo}}\sim10^{13.5}\,\rm{M_{\odot}}$ at $z=2$, respectively. For these four haloes, we do not implement black hole seeding or growth. In Figure \ref{fig:evolution_halos} we show the evolution of the halo mass of each of the eight haloes, as well as the evolution of the stellar mass, dust mass, and star formation rate of their central galaxies. \\
\indent The mass resolution of our A-series simulations is $3.3\times10^4\,\rm{M_{\odot}}$ for gas and star particles and $1.7\times10^5\,\rm{M_{\odot}}$ for dark matter particles (high resolution, hereafter HR). The mass resolution of our B-series simulations is $2.7\times10^5\,\rm{M_{\odot}}$ for gas and star particles and $1.4\times10^6\,\rm{M_{\odot}}$ for dark matter particles (standard resolution, hereafter SR). The mass resolution of our C-series simulations is $2.2\times10^6\,\rm{M_{\odot}}$ for gas and star particles and $1.1\times10^7\,\rm{M_{\odot}}$ for dark matter particles (low resolution, hereafter LR). We describe the resolution tests in Appendix \ref{sec:appendix_res_tests}, and show that even the lowest resolution simulations are converged. 

\begin{figure*} 
\centering
\includegraphics[scale=0.67]{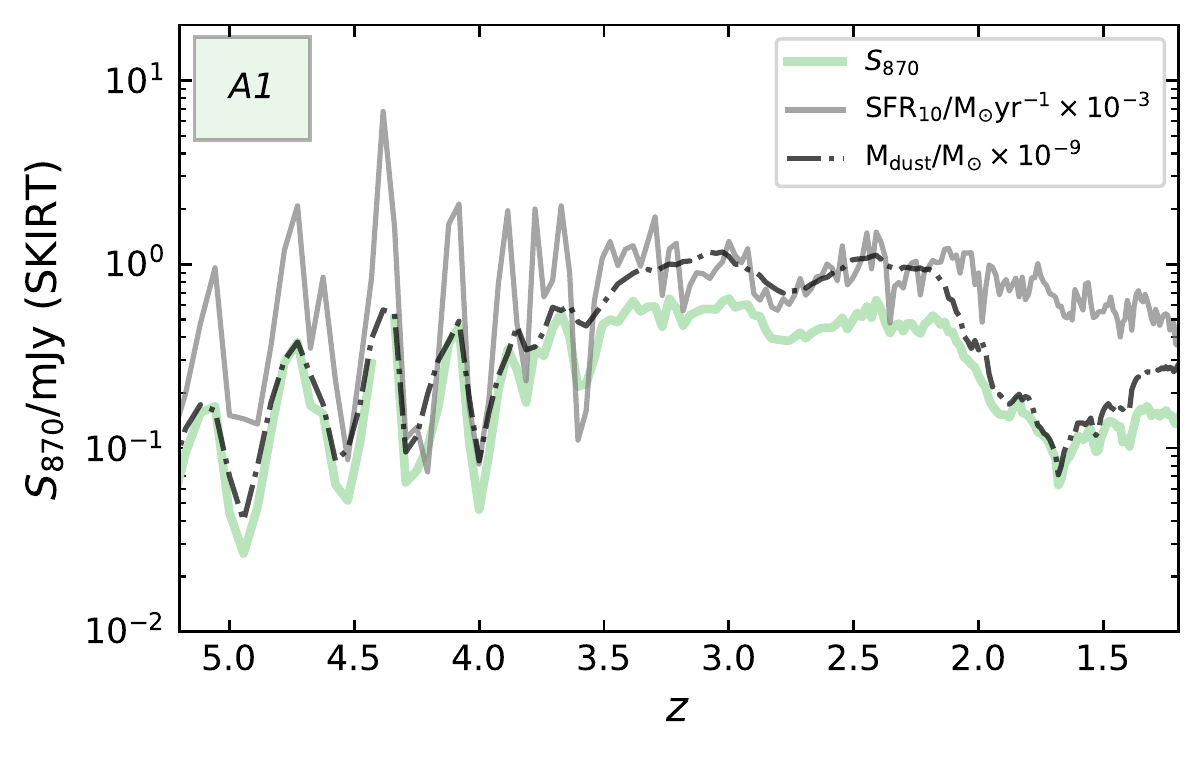}
\includegraphics[scale=0.67]{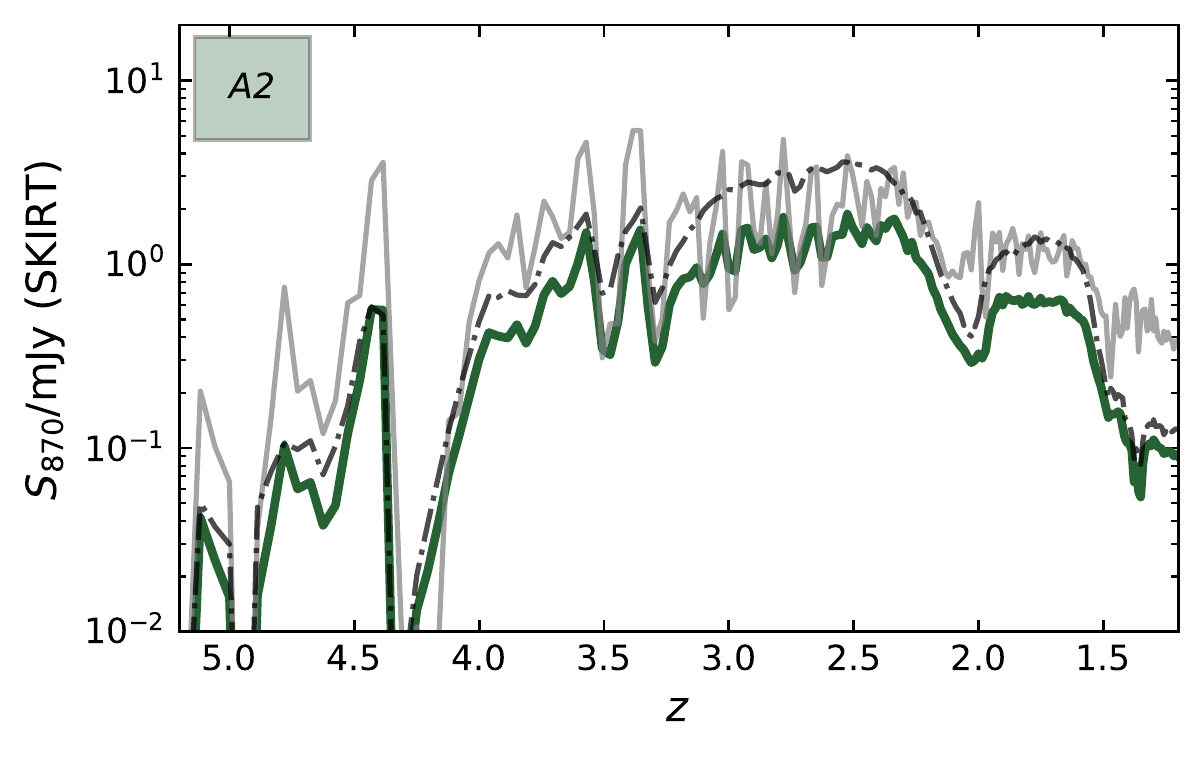}
\includegraphics[scale=0.67]{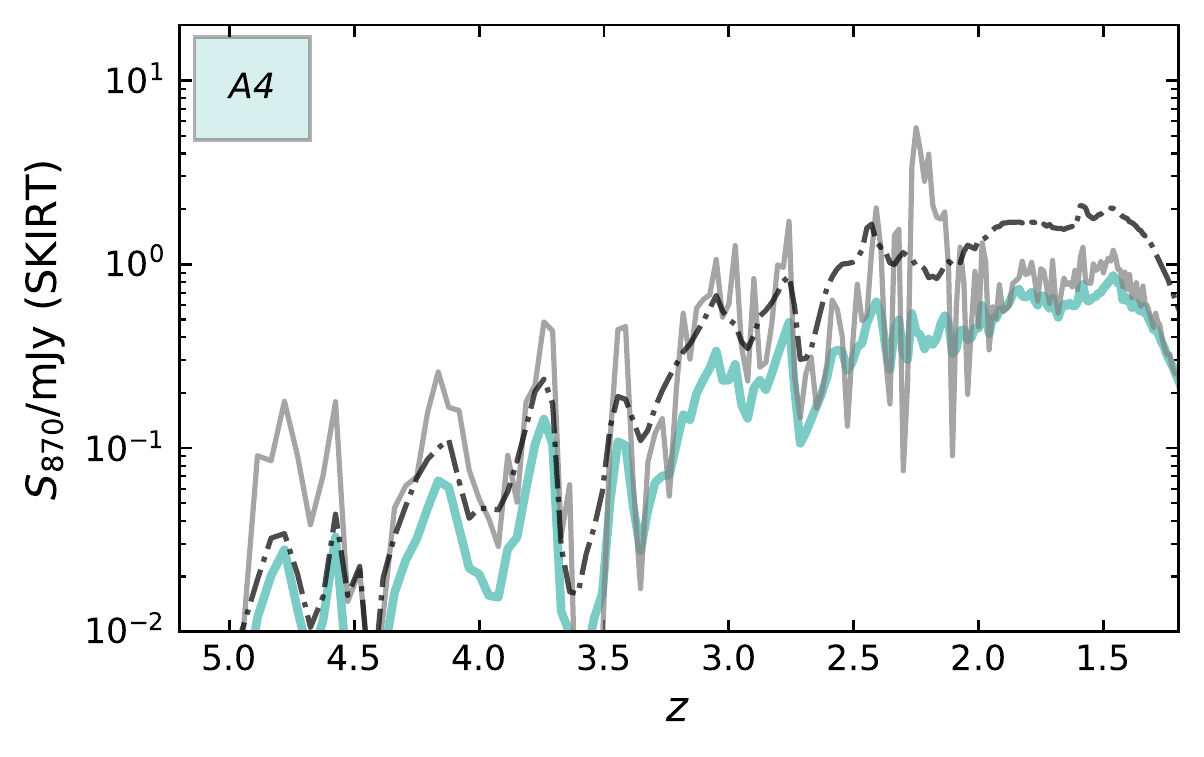}
\includegraphics[scale=0.67]{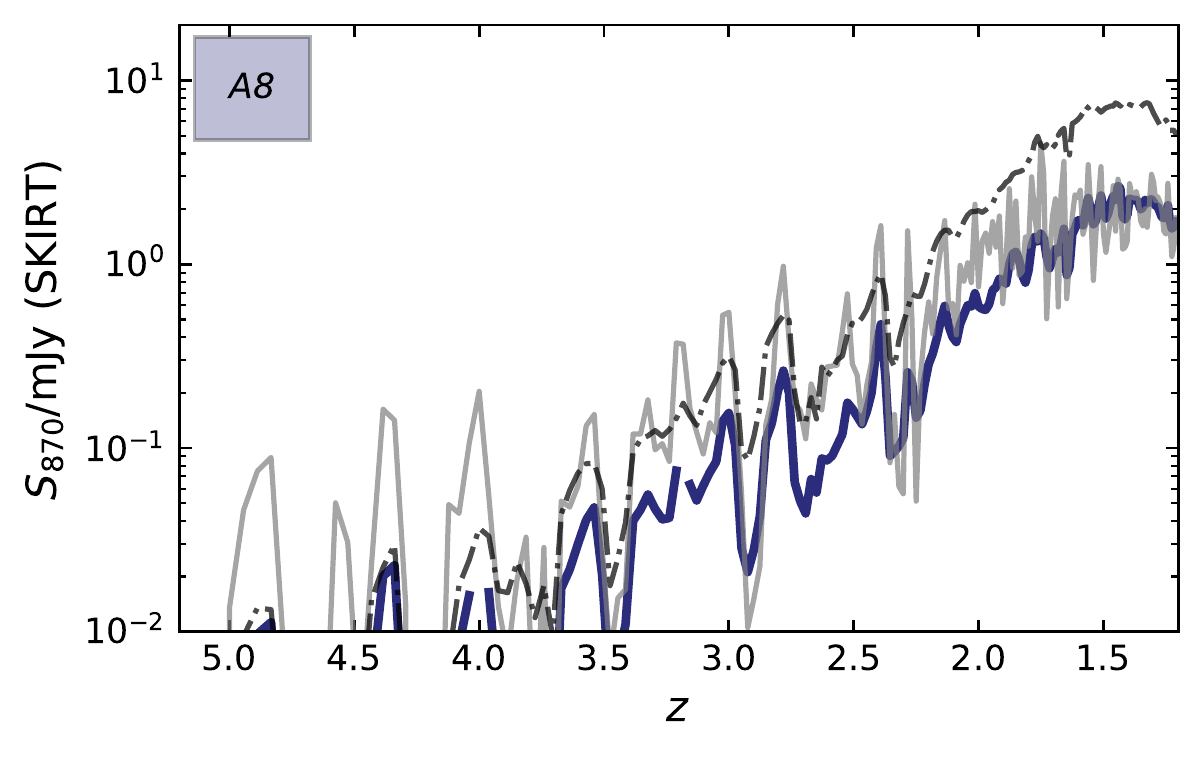}
\includegraphics[scale=0.67]{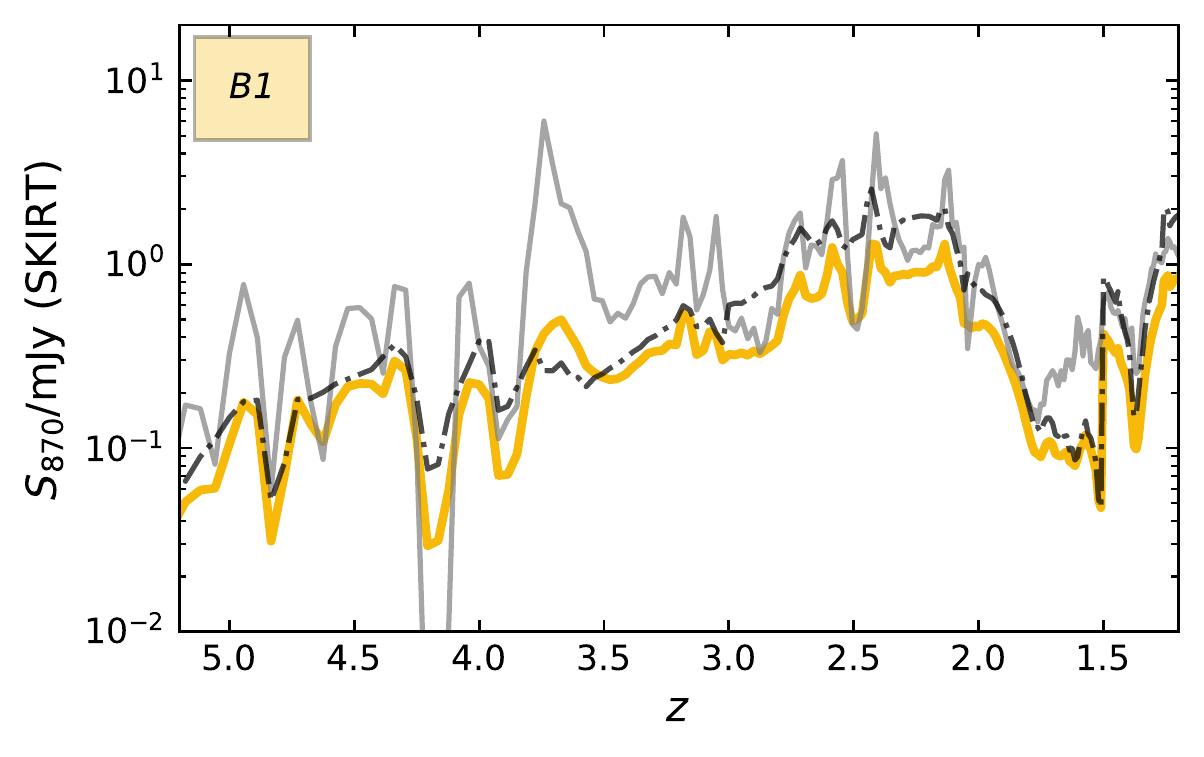}
\includegraphics[scale=0.67]{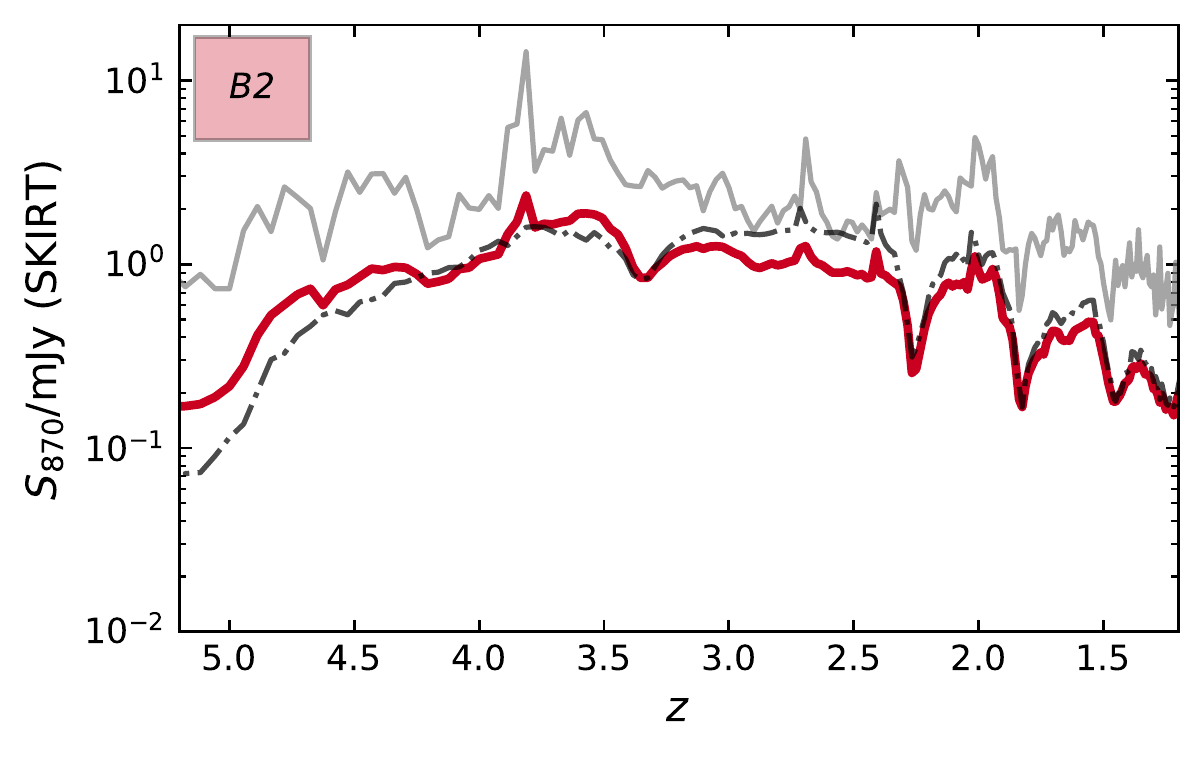}
\includegraphics[scale=0.67]{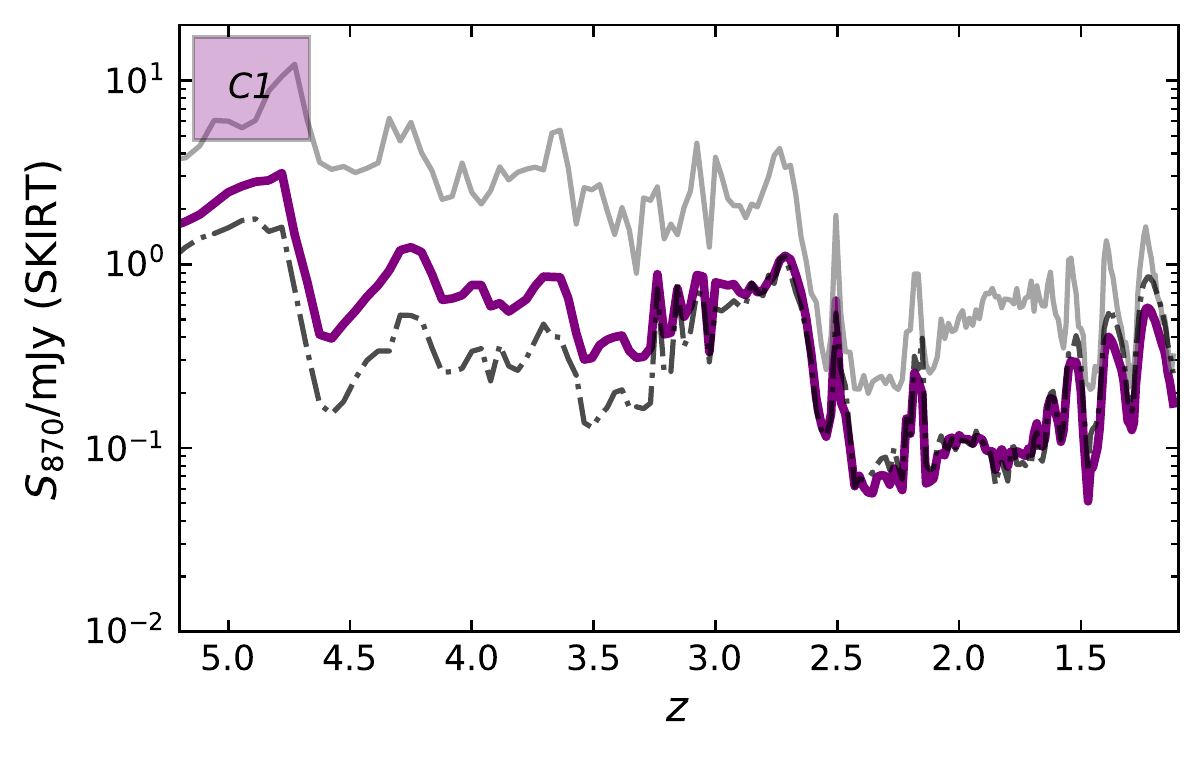}
\includegraphics[scale=0.67]{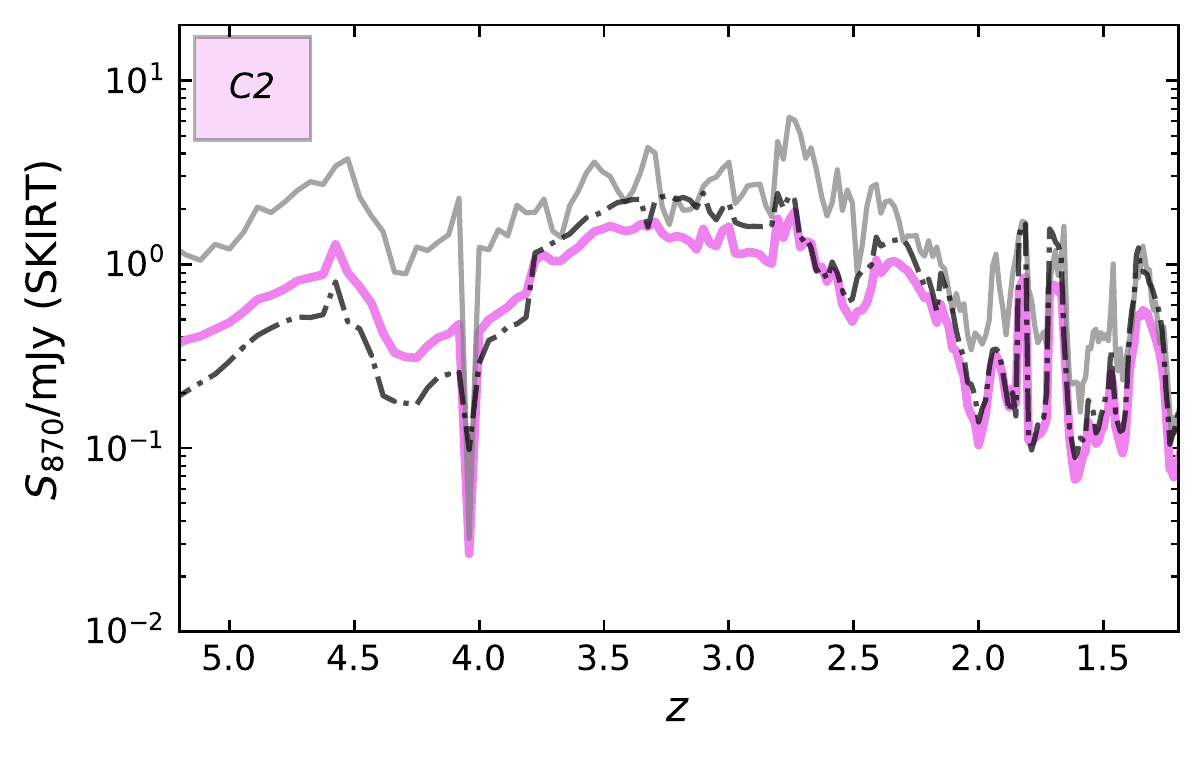}
\caption{The redshift evolution of star formation rate (averaged over $10\,\rm{Myr}$), dust mass and {\sc skirt}-predicted $870\,\mu\rm{m}$ flux density for the eight FIRE-2 central galaxies simulated, all calculated within 3D apertures of radius $0.1\,R_{\rm{vir}}$. The galaxies display a range of star formation histories and $870\,\mu\rm{m}$ flux densities, with the most sub-millimeter bright snapshots reaching $\sim3\,\rm{mJy}$. At early times, feedback from bursty star formation expels gas; the derived dust mass hence varies on relatively short timescales.}
\label{fig:A1_4_vs_z}
\end{figure*}

\begin{figure*} 
\centering
\includegraphics[scale=0.26]{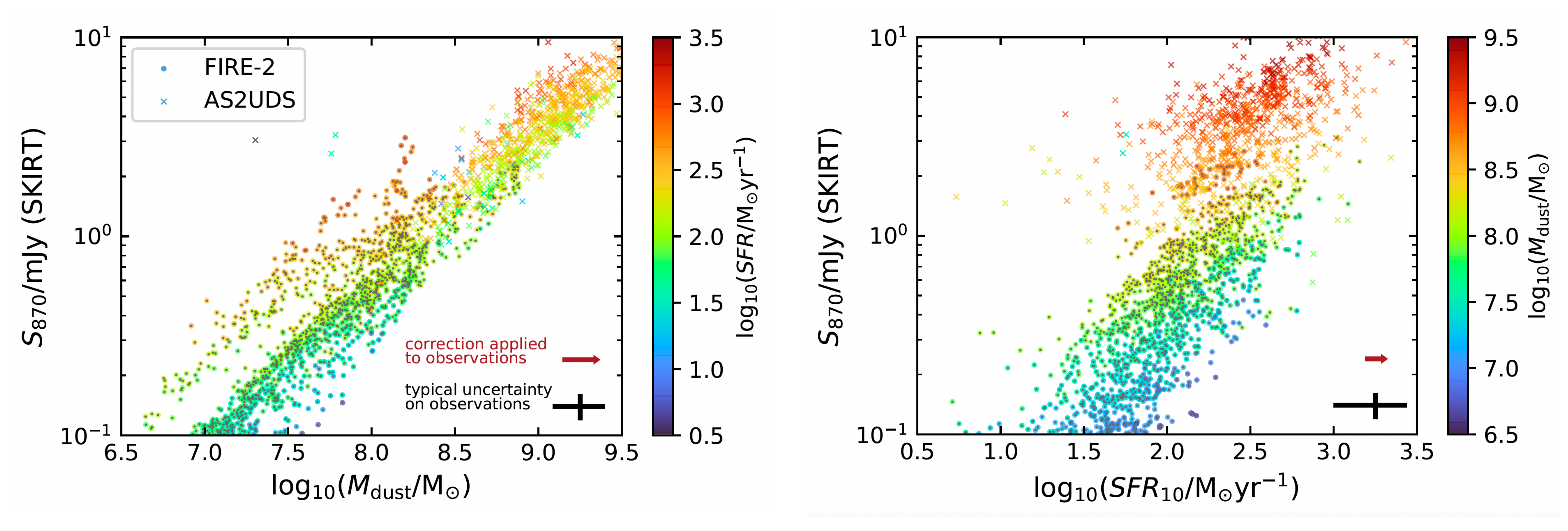}
\caption{Left: {\sc skirt}-predicted (observed-frame) $870\,\mu\rm{m}$ flux density versus dust mass, colour-coded by the $10\,\rm{Myr}$ average star formation rate of the eight FIRE-2 haloes. Right: {\sc skirt}-predicted (observed-frame) $870\,\mu\rm{m}$ flux density versus SFR, colour-coded by dust mass (we add small grey dots in the center of each of these symbols, to aid visibility). We also show observational data from the AS2UDS sample (primarily comprising $S_{870}>1\,\rm{mJy}$ sources; \citealt{Dudzeviciute2019}), on the same colour-scale. Dust masses and SFRs are derived for the AS2UDS sources using the {\sc{magphys}} SED fitting code, with small corrections applied to account for systematic bias. These are shown here with brown arrows: $\sim0.2\,\rm{dex}$ for $M_{\rm{dust}}$ and $\sim0.1\,\rm{dex}$ for $\rm{SFR}$, as derived by \protect\cite{Dudzeviciute2019}. Typical uncertainties for the observationally-derived parameters and flux densities are shown in black. The observational data are broadly consistent with the predictions based on applying {\sc{skirt}} to our simulated galaxies. It is clear that $S_{870}$ depends on both dust mass and star formation rate. At fixed dust mass, galaxies with higher star formation rates have higher $S_{870}$ . At fixed star formation rate, galaxies with larger dust masses also have higher $S_{870}$ .}
\label{fig:obs_flux_sfr_dust}
\end{figure*}

\subsection{Performing radiative transfer to predict realistic galaxy fluxes}\label{sec:skirt_details}
Modelling the multi-wavelength emission from simulated galaxies can be challenging due to the need to account for the complex three-dimensional spatial distribution of radiation sources (stars and AGN), and the dust particles that scatter, absorb and re-emit light. Performing the detailed radiative transfer modelling necessary can be computationally expensive \citep[see][for a review]{Steinacker2013}. \\
\indent In this work we follow the methods described in \cite{Cochrane2019} to apply the Stellar Kinematics Including Radiative Transfer ({\sc skirt})\footnote{\url{http://www.skirt.ugent.be}} Monte Carlo radiative transfer code \citep{Baes2011,Camps2014} to FIRE-2 galaxies in post-processing. Monte Carlo radiative transfer codes like {\sc skirt} treat the radiation field as a flow of photons through the dusty medium of a galaxy to compute the effects of dust absorption, scattering, and re-emission of the absorbed light, including dust self-absorption. While \cite{Cochrane2019} selected only the most sub-mm bright snapshots for their analysis, we have run {\sc skirt} on all simulation snapshots for each of the eight haloes studied. We briefly summarise the method here.\\
\indent Gas and star particles (along with their properties such as age and metallicity) are drawn from the FIRE-2 simulations at each snapshot. Prior to input to {\sc skirt}, we rotate the particles to align the line-of-sight with the angular momentum vector of the gas particles within $0.1R_{\rm{vir}}$, such that the galaxy is `face on' (for disk galaxies) at a viewing angle of 0 degrees. Dust particles are assumed to follow the distribution of the metals, with a dust-to-metals mass ratio of $0.4$ \citep{Dwek1998,James2002}. We assume dust destruction at $>10^6\,{\rm K}$ \citep{Draine1979,Tielens1994}. We model a mixture of graphite, silicate and PAH grains using the \cite{Weingartner2001} Milky Way dust prescription. Star particles are assigned \citet{Charlot2003} SEDs based on their ages and metallicities, assuming a \cite{Chabrier2003} IMF. We use an octree dust grid, in which cell sizes are adjusted according to the dust density distribution, with the condition that no dust cell may contain more than $0.0001\%$ of the total dust mass of the galaxy. We use $10^{6}$ photon packages, following tests that show convergence see Appendix \ref{sec:appendix_skirt_tests}). \\
\indent The {\sc skirt} output at a given modelled source orientation comprises spatially-resolved images (on a grid with pixel side length $25\,\rm{pc}$) at $\sim100$ discrete rest-frame wavelengths, spaced uniformly in log(wavelength) between rest-frame UV and FIR wavelengths. We also add the rest-frame wavelength corresponding to each ALMA band, for each snapshot (i.e. different rest-frame wavelengths that depend on the redshift of the snapshot). We model seven detectors at different orientations with respect to the disk plane of the galaxy. The field of view is set to 10\% of the virial radius for each galaxy snapshot studied. Tests confirm that this radius encloses all the sub-millimeter flux density from the simulated galaxy, even at the highest redshifts studied, where the virial radius is smallest.\\
\indent Although {\sc skirt} provides predictions for UV-FIR fluxes, in this paper, we focus on observed-frame sub-mm emission. The measured emission at these wavelengths is insensitive to orientation since we are usually in the optically thin regime \citep[e.g.][]{Cochrane2019}, so we consider just the `face-on' {\sc{skirt}}-generated SED. In Figure \ref{fig:A1_4_vs_z}, we plot the evolution of the predicted (observed-frame) $870\,\mu\rm{m}$ flux density, $S_{870}$, SFR and dust mass, for each of the eight simulated haloes. All physical quantities are calculated within a 3D aperture of radius $0.1\,R_{\rm{vir}}$.

\section{Scaling relations between $\mathbf{S_{870}}$, SFR and dust mass}\label{sec:scaling_relations}
In this section, we derive a simple power-law scaling relation between SFR, dust mass and $870\,\mu\rm{m}$ flux density. First, we establish that the FIRE-2 galaxies broadly follow the same trends in $\rm{SFR}-S_{870}$ and $M_{\rm{dust}}-S_{870}$ as observed galaxies. In Figure \ref{fig:obs_flux_sfr_dust}, we plot dust mass and SFR against {\sc skirt}-predicted $S_{870}$ for all snapshots studied for each of the eight haloes. We overplot observational data from the AS2UDS sample of $\sim700$ sub-mm sources \cite{Dudzeviciute2019}, homogeneously selected from the SCUBA-2 Cosmology Legacy Survey $850\,\mu\rm{m}$ map of the UKIDSS/UDS field \citep{Stach2019b}. \cite{Dudzeviciute2019} use the {\sc magphys} SED fitting code to infer dust mass and SFR from UV-radio photometry. \cite{Dudzeviciute2019} test the reliability of the {\sc magphys}-derived quantities by fitting mock photometry of simulated EAGLE galaxies with known physical properties. They thereby infer correction factors (`true' dust mass is a factor of $1/0.65$ higher than derived by {\sc magphys}, while `true' SFR is a factor of $1/0.8$ higher). We apply these correction factors to their derived estimates. From these plots, it is clear that our predictions are in reasonable agreement with the observationally derived relations, though we do not reach the brightest sub-mm flux densities observed in AS2UDS, due to the limited parameter space probed by our small set of simulations (in particular, we do not reach extremely high values of SFR or $M_{\rm{dust}}$). \\
\indent As seen from Figure \ref{fig:obs_flux_sfr_dust}, it is clear from both observations and our simulations that both dust mass and SFR correlate with $S_{870}$. Neither dust mass nor SFR predicts $S_{870}$ alone: at fixed dust mass, galaxies with higher star formation rates have higher $S_{870}$, and at fixed star formation rate, galaxies with larger dust masses also have higher $S_{870}$. In Section \ref{sec:flux_predictions_3param}, we fit a power-law relation to relate the three quantities.

\subsection{Predicting $\mathbf{S_{870}}$ from SFR and dust mass for FIRE galaxies}\label{sec:flux_predictions_3param}
\begin{figure} 
\centering
\includegraphics[scale=0.46]{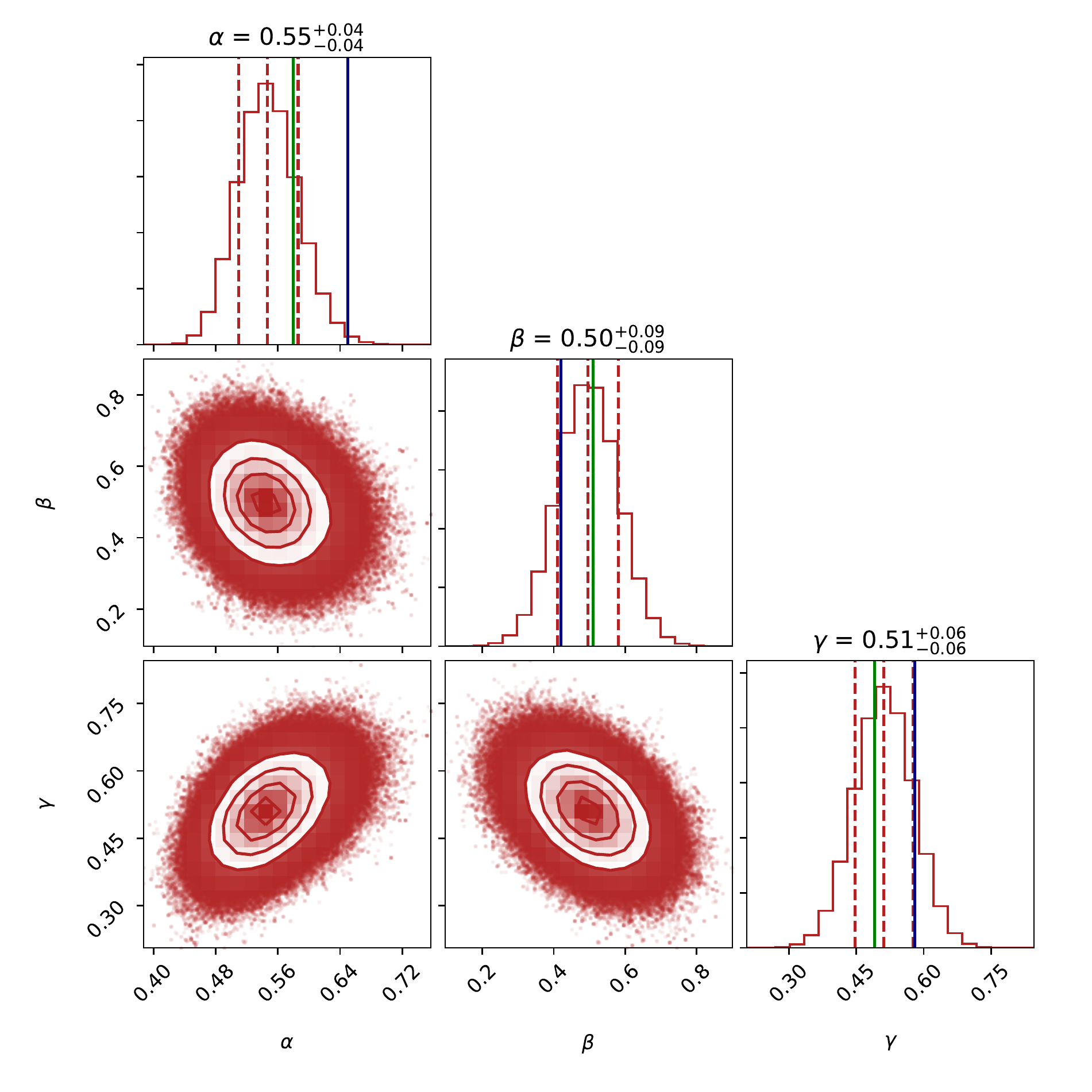}
\caption{The posterior distribution of the parameters $\alpha$, $\beta$ and $\gamma$, as described in Equation \ref{eq:eq1}, fitted to the FIRE-2 snapshots (using a {\it{corner}} plot; \protect\citealt{Foreman-Mackey2016}). Parameter estimates from \protect\cite{Hayward2011} and \protect\cite{Lovell2020} are overplotted in blue and green, respectively. Our fits show clear degeneracies between the fitted parameters, which may help understand discrepancies among previous fits.}
\label{fig:triangle_plot_FIRE}
\end{figure}

\begin{figure} 
\centering
\includegraphics[scale=0.615]{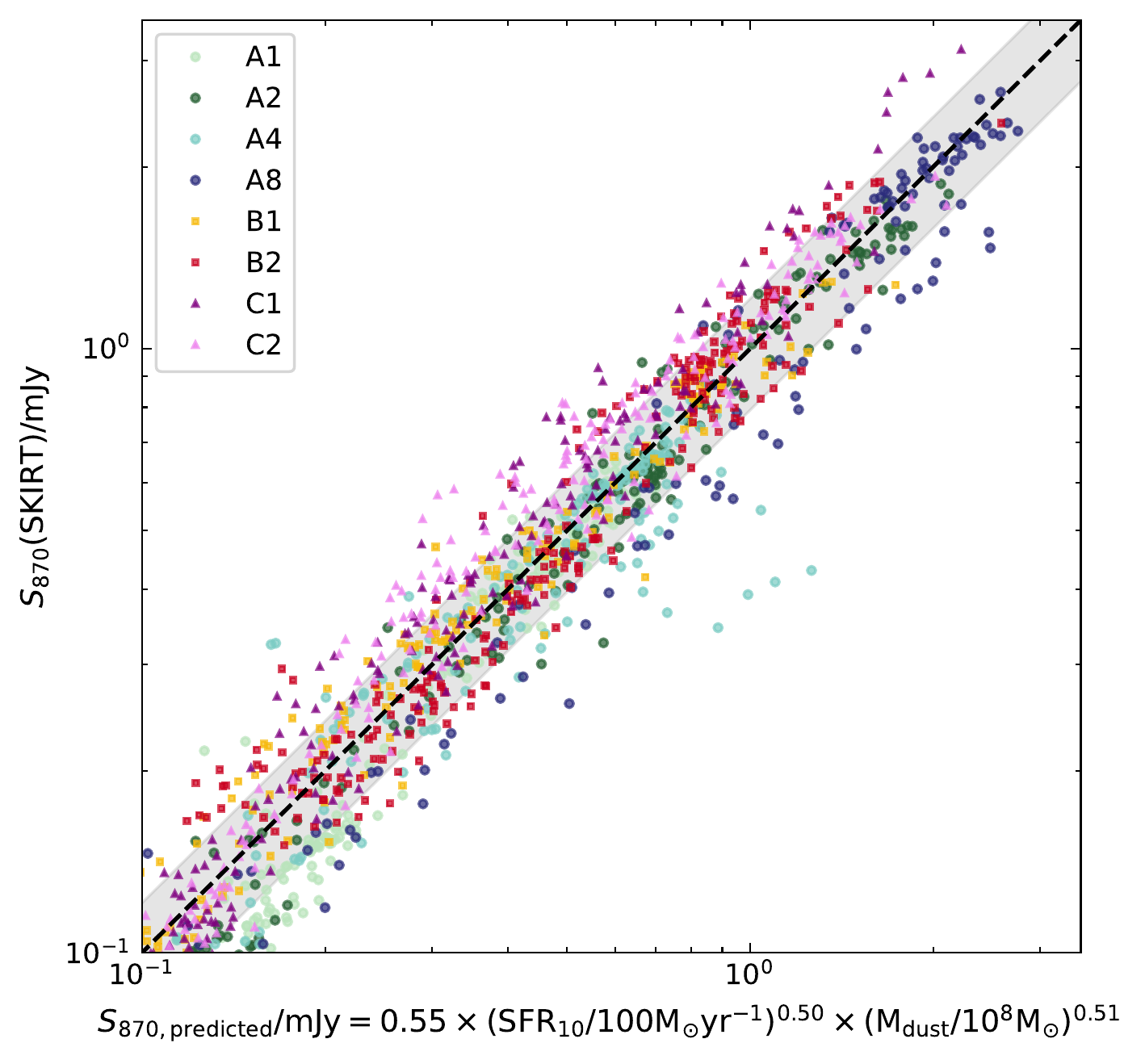}	
\caption{The {\sc{skirt}}-predicted $870\,\mu\rm{m}$ flux density versus the value predicted by $\rm{SFR}_{10}$ and dust mass using our fitted formula, for all snapshots with $S_{870}>0.1\,\rm{mJy}$, colour-coded by halo. For snapshots where $S_{870}>0.1\,\rm{mJy}$, the mean $S_{870,\rm{SKIRT}}/S_{870,\rm{formula}}$ is $1.017$. The standard deviation in $S_{870,\rm{SKIRT}}/S_{870,\rm{formula}}$ (plotted as the grey shaded region) is $0.21$. This indicates that $S_{870\,\mu\rm{m}}$ can typically be predicted to within $0.09\,\rm{dex}$ from the star formation rate and dust mass of a simulated galaxy alone.}
\label{fig:fitted_relation}
\end{figure}

Following \cite{Hayward2011}, we fit the following relation using the {\it{emcee}} package \citep{Foreman-Mackey2013}:
\begin{equation}\label{eq:eq1}
    S_{870}/{\rm{mJy}} = \alpha \Bigg(\frac{{\rm{SFR}}_{10}}{100\,{\rm{M_{\odot}}}{\rm{yr}}^{-1}}\Bigg)^{\beta} \Bigg(\frac{M_{\rm{dust}}}{10^{8}\,\rm{M_{\odot}}}\Bigg)^{\gamma}.
\end{equation}
Here, $\rm{SFR}_{10}$ corresponds to the star formation rate averaged over the $10\,\rm{Myr}$ before the snapshot; this approximately corresponds to the timescales probed by observable SFR tracers \citep[e.g.][]{FloresVelazquez2021}. $M_{\rm{dust}}$ is the total dust mass used by {\sc{skirt}} in the radiative transfer. We were motivated to adopt this functional form due to its success in fitting radiative transfer-predicted flux densities previously \citep{Hayward2011,Hayward2013b} and its simplicity.\\
\indent We present our results in Figure \ref{fig:triangle_plot_FIRE}. The derived parameters are $\alpha=0.55\pm0.04$, $\beta=0.50\pm0.09$, and $\gamma=0.51\pm0.06$. \cite{Hayward2011} derived $\alpha=0.65$, $\beta=0.42$, and $\gamma=0.58$. We overplot their results in blue. \cite{Lovell2020} perform a similar fit to galaxies in S{\sc IMBA}, and obtain $\alpha=0.58$, $\beta=0.51$, and $\gamma=0.49$. We overplot their results in green. No uncertainties are provided by either of these works. While the estimates of \cite{Lovell2020} are within $1\,\sigma$ of ours, those of \cite{Hayward2011} deviate more. The correlations between the fitted parameters shown in Figure \ref{fig:triangle_plot_FIRE} help explain some discrepancies. While their estimate of $\beta$ is $\sim1\,\sigma$ lower than ours, their estimate of $\gamma$ is $\sim1\,\sigma$ greater. This makes sense given the negative correlations between the two parameters shown in \ref{fig:triangle_plot_FIRE}. Their value of $\alpha$ deviates from ours by over $2\,\sigma$. Despite these differences, it is encouraging that the derived relations agree to the degree that they do; \cite{Hayward2011}, for example, used highly idealized simulations with less sophisticated models for star formation, the ISM and stellar feedback.\\
\indent In Figure \ref{fig:fitted_relation}, we compare the sub-mm flux densities predicted by our formula to the flux densities derived using the radiative transfer, for the FIRE galaxy snapshots used in deriving the scaling relation. This shows how well the combination of $\rm{SFR}_{10}$ and dust mass can predict sub-mm flux density. For galaxies with $S_{870}>0.1\,\rm{mJy}$, the mean $S_{870,\rm{SKIRT}}/S_{870,\rm{formula}}$ is $1.017$. The standard deviation is $0.21$. This indicates that $S_{870}$ can typically be predicted to within $0.09\,\rm{dex}$ from the star formation rate and dust mass of a simulated galaxy alone. The tightness of this relation is encouraging. We explore the validity of application to other simulations in our application of FIRE-derived scaling relations to EAGLE galaxies in Section \ref{sec:eagle_application}.

\subsection{Comparison with observational data}
\begin{figure} 
    \centering
    \includegraphics[scale=0.68]{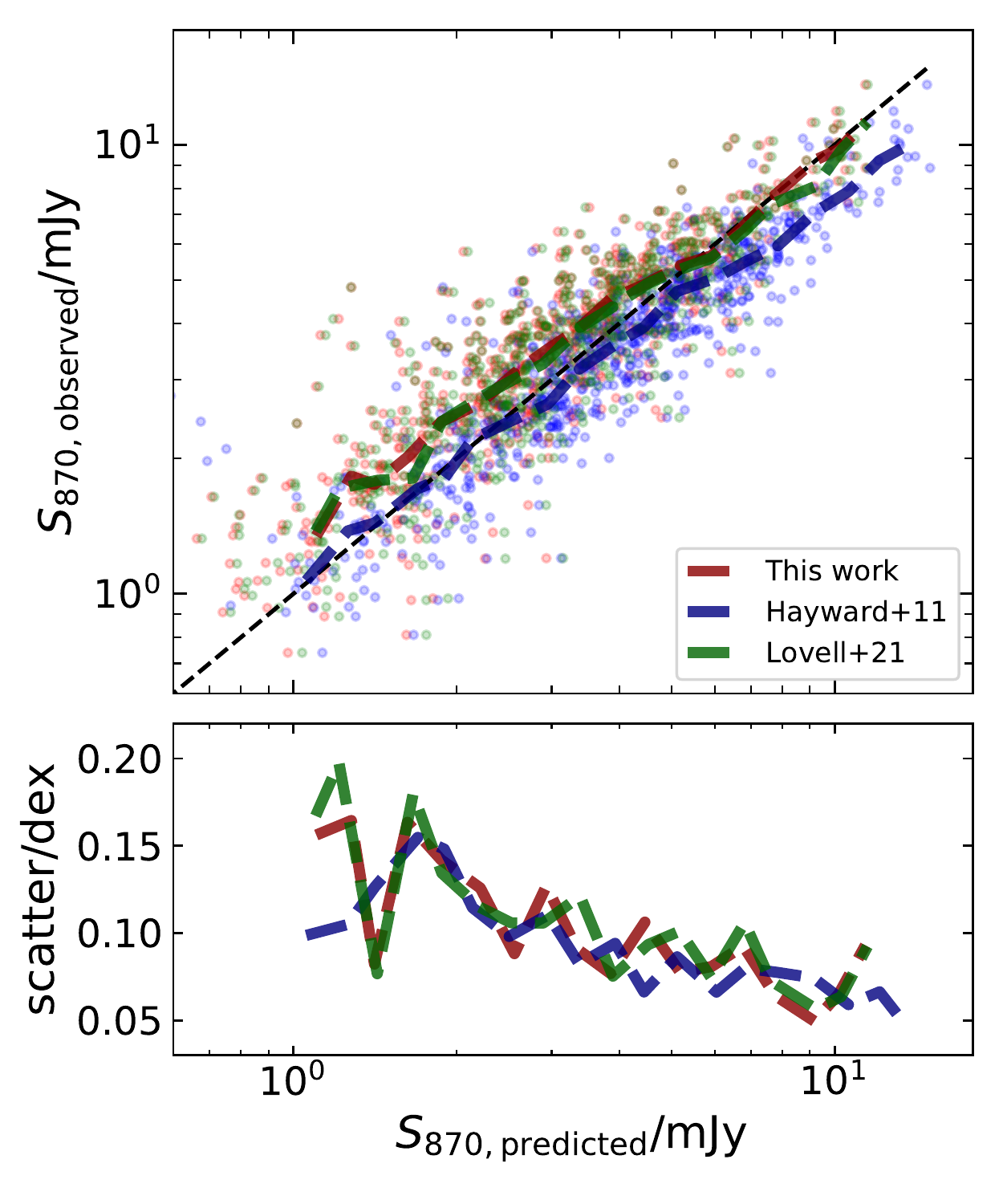}
	\caption{Upper panel: $870\,\mu\rm{m}$ flux density observed by ALMA against that predicted using dust mass and SFR ({\sc{magphys}} estimates) for the AS2UDS galaxies. The flux densities predicted from the {\sc{magphys}-}derived dust mass and SFR using our FIRE-derived scaling relation are in good agreement with those observed by ALMA, with median $S_{870,\rm{observed}}/S_{870,\rm{formula}}=1.22$. Lower panel: the standard deviation in observed-to-predicted flux densities. The dashed lines on both panels show running medians (using 20 bins, log-spaced between $0$ and $15\,\rm{mJy}$), highlighting both the low scatter and the lack of strong trends with flux density.}
    \label{fig:dudz_comparison}
\end{figure}

To validate our power law relation, we apply it to the (systematic bias-corrected) dust mass and SFR values predicted by {\sc{magphys}} for the AS2UDS galaxies. In Figure \ref{fig:dudz_comparison} (upper panel), we show observed versus predicted flux densities. The median value of $S_{870,\rm{observed}}/S_{870,\rm{formula}}$ is $1.22$ (see middle panel). The standard deviation in $S_{870,\rm{observed}}/S_{870,\rm{formula}}$ is shown in the lower panel and is fairly constant with observed flux density, remaining at $\sim0.1\,\rm{dex}$. This adds confidence that our simulation-derived relations are consistent with those of real observed galaxies.Note that without the correction for bias in the {\sc{magphys}-}inferred physical properties, the $S_{870,\rm{observed}}$ is systematically higher than predicted by the relation by $\sim0.2\,\rm{dex}$.

\section{Extension to other FIR/sub-mm flux densities}\label{sec:four_flux_densities}
\begin{figure*} 
    \hspace{-0.3cm}
	\includegraphics[scale=0.48]{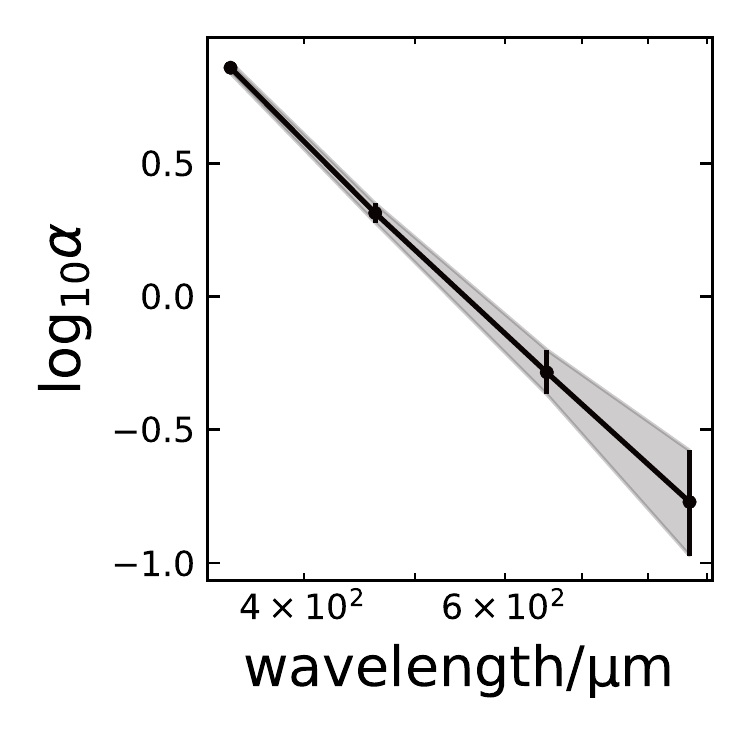}
	\hspace{-0.3cm}
	\includegraphics[scale=0.48]{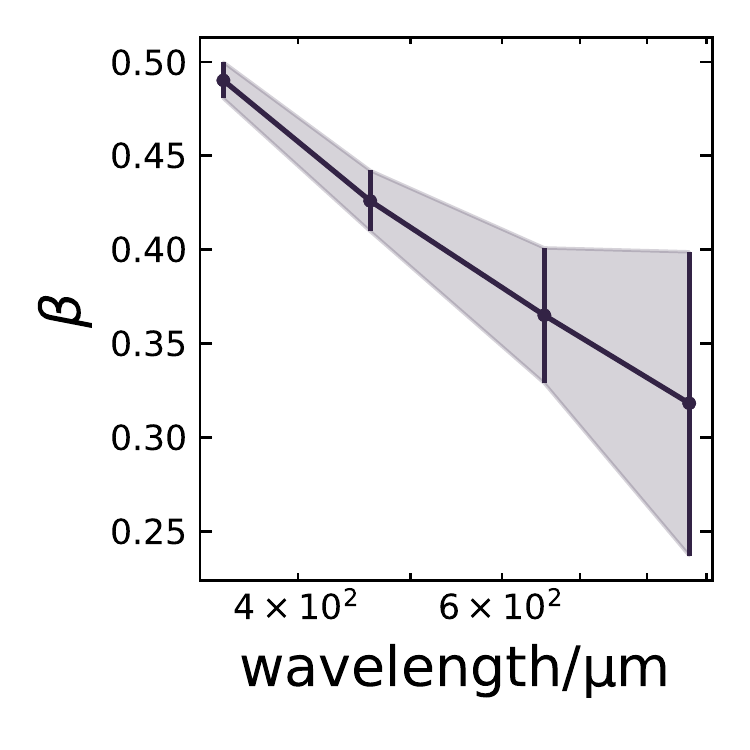}
	\hspace{-0.3cm}
	\includegraphics[scale=0.48]{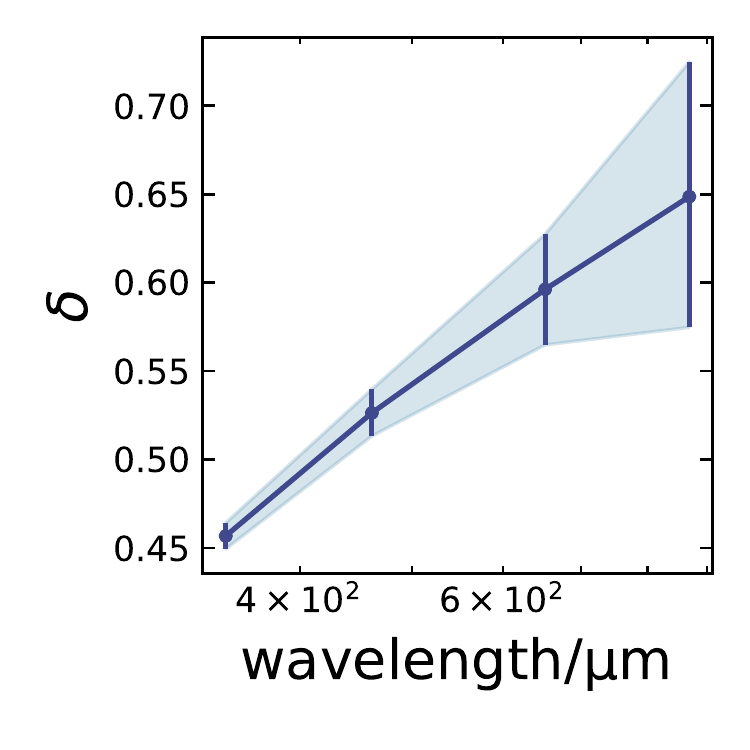}
	\hspace{-0.25cm}
	\includegraphics[scale=0.48]{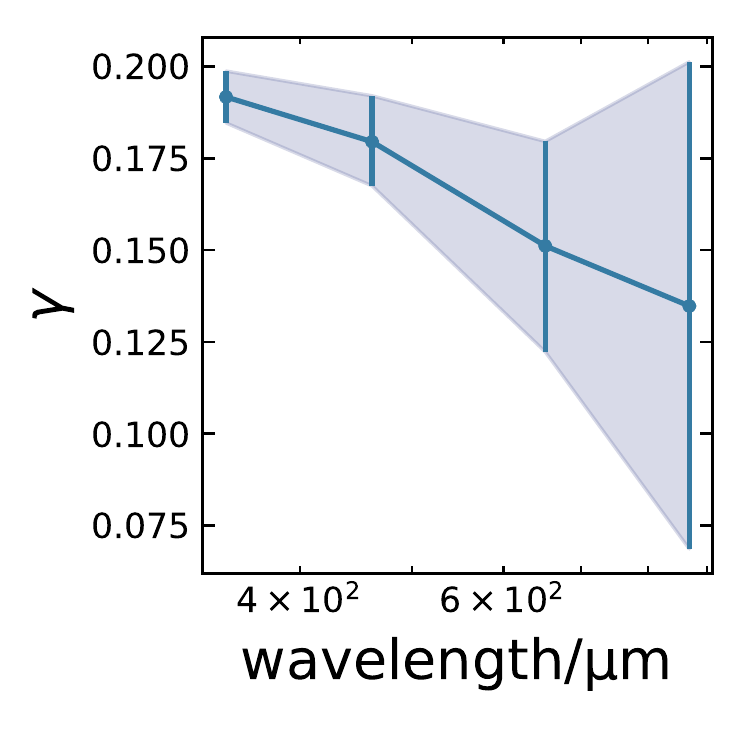}
	\hspace{-0.32cm}
	\includegraphics[scale=0.48]{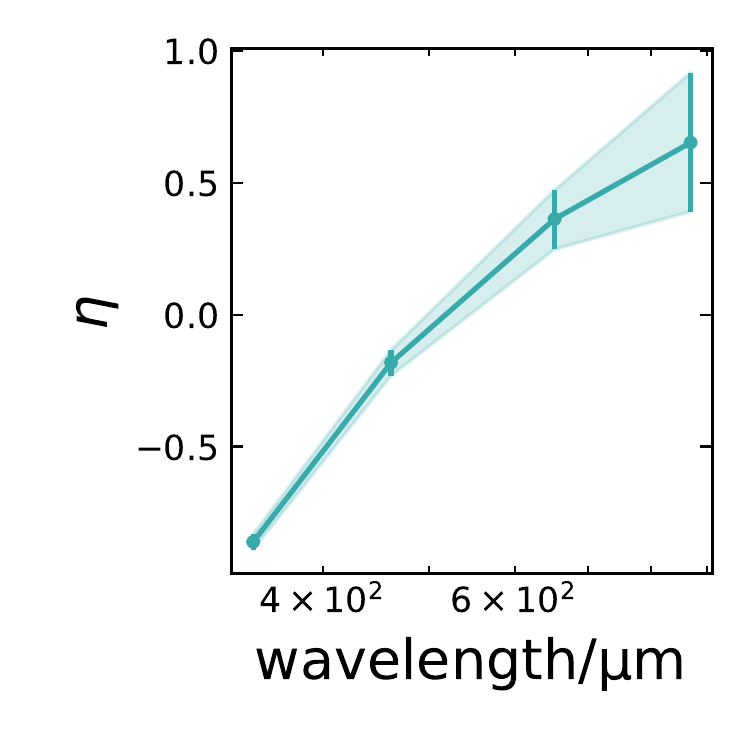}
	\caption{Parameters of $S_{\nu}/{\rm{mJy}} = \alpha \Bigg(\frac{\rm{SFR}_{10}}{100\,\rm{M_{\odot}yr^{-1}}}\Bigg)^{\beta} \Bigg(\frac{M_{\rm{dust}}}{10^{8}\,\rm{M_{\odot}}}\Bigg)^{\delta}\Bigg(\frac{M_{\rm{\star}}}{10^{10}\,\rm{M_{\odot}}}\Bigg)^{\gamma}(1+z)^{\eta}$ fitted using FIRE-2 snapshots at $1<z<4$. The estimate for each parameter is the median of its posterior distribution, with $1\,\sigma$ uncertainties calculated using the $16^{\rm{th}}$ and $84^{\rm{th}}$ percentiles. We use the $10\,\rm{Myr}$-averaged SFR. Notably, $\beta$ decreases with increasing wavelength, reflecting the increasing importance of SFR in determining sub-mm flux density closer to the peak of the dust SED. $\delta$ displays the opposite trend, reflecting the increasing dominance of dust mass in driving sub-mm flux density at longer wavelengths.}
    \label{fig:params_vs_wavelength}
\end{figure*}

\begin{figure} 
    \centering
    \includegraphics[scale=0.65]{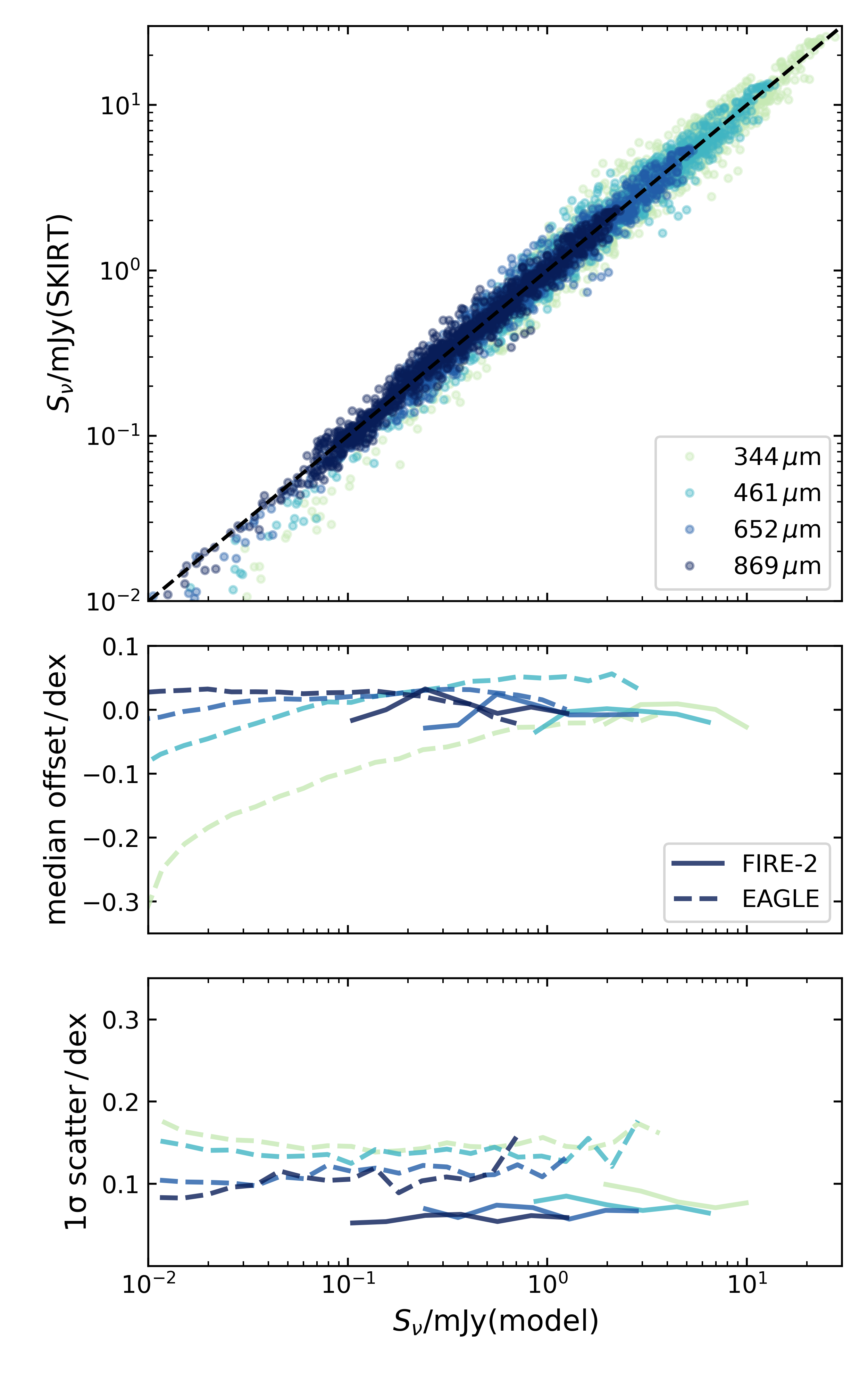}
	\caption{Top panel: {\sc{skirt}}-predicted versus model-predicted flux density for ALMA Bands 7, 8, 9 \& 10, for the FIRE-2 snapshots. Middle panel: running median offset from the model-predicted flux density for the FIRE-2 galaxies (solid lines) and EAGLE galaxies (dashed lines). Bottom panel: running scatter in the relation shown in the upper panel. The flux densities predicted for EAGLE galaxies using our scaling relations are typically within $0.2\,\rm{dex}$ of the {\sc{skirt}}-predicted values.}
    \label{fig:all_bands_calibration}
\end{figure}

Following the success of our simple power law scaling between star formation rate, dust mass and $870\,\mu\rm{m}$ flux density, presented in Section \ref{sec:scaling_relations}, we attempt to predict the FIR SED at more wavelengths via a similar parametrisation. We develop a model that may be applied to the global properties of simulated galaxies to make rapid predictions for galaxy number counts at various sub-mm wavelengths. These simulated galaxies could include large box simulations, for which running radiative transfer would be computationally infeasible; simulations with lower resolution (and more poorly resolved ISM) for which the output of radiative transfer may be unreliable; and semi-analytic models, where detailed spatial distributions of stars and dust do not exist. \\
\indent We fit the following relation to SKIRT-predicted observed-frame flux densities at $345\,\mu\rm{m}$, $462\,\mu\rm{m}$, $652\,\mu\rm{m}$ and $870\,\mu\rm{m}$ (corresponding to ALMA Bands 10, 9, 8 \& 7, respectively), for snapshots in the redshift range $1<z<4$:

\begin{equation}\label{eq:eq2}
    S_{\nu}/{\rm{mJy}} = \alpha \Bigg(\frac{{\rm{SFR}}_{10}}{100\,{\rm{M_{\odot}yr}^{-1}}}\Bigg)^{\beta}\Bigg(\frac{M_{\star}}{10^{10}\,\rm{M_{\odot}}}\Bigg)^{\gamma} \Bigg(\frac{M_{\rm{dust}}}{10^{8}\,\rm{M_{\odot}}}\Bigg)^{\delta}(1+z)^{\eta}.
\end{equation}
Redshift is an important added parameter in this extension to other wavelengths, since we are fitting various observed-frame wavelengths and hence cannot rely on the negative $K-$correction for redshift-independence. As in Section \ref{sec:flux_predictions_3param}, we use the SFR averaged over the $10\,\rm{Myr}$ prior to each snapshot. We again use the {\it{emcee}} package. Our best-fitting parameters are presented in Table \ref{Table:fitted_params_10Myr} and plotted as a function of wavelength in Figure \ref{fig:params_vs_wavelength}. We note the following trends in fitted parameters with wavelength. The normalisation factor, $\alpha$, decreases with increasing wavelength, in line with the expected decrease in flux density with wavelength. $\beta$ decreases with increasing wavelength, reflecting the increased role of SFR in determining sub-mm flux density around the peak of the dust SED. $\delta$ increases with wavelength, reflecting the increasing dominance of dust mass in driving sub-mm flux density as one approaches the Rayleigh-Jeans tail (see also \citealt{Cochrane2022a}). $\gamma$ is low ($\sim0.1-0.2\,\rm{dex}$) at all wavelengths, showing the weak dependence of sub-mm flux density on stellar mass. $\eta$ increases with wavelength, in line with expectations from the negative $K$-correction.\\
\indent Our derived scaling relations recover the {\sc{skirt}}-derived flux densities accurately at all four wavelengths. Median values of offsets and scatter are $[-0.008,0.11]\,\rm{dex}$, $[-0.012,0.087]\,\rm{dex}$, $[-0.009,0.071]\,\rm{dex}$ and $[0.0011,0.061]\,\rm{dex}$, for $345\,\mu\rm{m}$, $462\,\mu\rm{m}$, $652\,\mu\rm{m}$ and $870\,\mu\rm{m}$, respectively. Using our scaling relations, the majority of the flux densities are predicted to within $0.1\,\rm{dex}$, with no strong trends with flux density (see Figure \ref{fig:all_bands_calibration}). In Section \ref{sec:eagle_application}, we apply the derived relations to simulated galaxies from EAGLE.

\begin{table*}
\begin{center}
\begin{tabular}{c|c|c|c|c|c}
$\lambda/\mu\rm{m}$ & $\log_{10}\alpha$ & $\beta$ &  $\delta$ & $\gamma$ &  $\eta$ \\ 
\hline
$345$ & $0.86\pm0.02$ & $0.49\pm0.01$ & $0.46\pm0.02$ & $0.19\pm0.01$ & $-0.86\pm0.03$ \vspace{0.15cm}\\ 
$462$ & $0.31\pm0.03$ & $0.43^{+0.01}_{-0.02}$ & $0.53\pm0.02$ & $0.18\pm0.01$ & $-0.18\pm0.05$ \vspace{0.15cm}\\ 
$652$ & $-0.29^{+0.09}_{-0.08}$ & $0.36\pm0.04$ & $0.60^{+0.03}_{-0.04}$ & $0.15\pm0.03$ & $0.36\pm0.11$ \vspace{0.15cm}\\ 
$870$ & $-0.77\pm0.2$ & $0.32\pm0.08$ & $0.65\pm0.07$ & $0.13\pm0.07$ & $0.65\pm0.25$
\end{tabular}
\caption{Parameters in Equation \ref{eq:eq2} fitted using FIRE-2 snapshots at $1<z<4$. The estimate for each parameter is the median of its posterior distribution, with $1\,\sigma$ uncertainties calculated using the $16^{\rm{th}}$ and $84^{\rm{th}}$ percentiles.}
\label{Table:fitted_params_10Myr}
\end{center}
\end{table*}

\section{Application of the FIRE-trained relations to the EAGLE simulation}\label{sec:eagle_application}
\subsection{The EAGLE simulation}\label{sec:eagle_sims}
The Virgo Consortium's Evolution and Assembly of GaLaxies and their Environments project (EAGLE\footnote{\url{http://icc.dur.ac.uk/Eagle/}}; \citealt{Crain2015,Schaye2015}) comprises a suite of $\Lambda\rm{CDM}$ simulations based on a modified version of the {\small GADGET 3} SPH code \citep{Springel2005}. The simulations use subgrid models for radiative cooling, star formation, stellar mass loss and metal enrichment, gas accretion onto black holes, black hole mergers, and energy feedback from both massive stars and AGN.
Feedback models are calibrated to the observationally inferred $z=0.1$ galaxy stellar mass function, the galaxy stellar mass - size relation, and the galaxy-black hole mass relation \citep{Crain2015}. EAGLE reproduces a number of other observed galaxy relations (among them, distributions of specific star formation rate, passive fractions, the evolution of the stellar mass function, galaxy rotation curves, and the Tully-Fisher relation), despite not being explicitly calibrated to match them \citep{Furlong2015,Schaller2015,Schaye2015,Trayford2015}.\\
\indent The various simulations within the EAGLE suite are presented in \cite{McAlpine2015}. We draw our galaxy samples from the largest box (version Ref-L100N1504; comoving side length $100\,\rm{Mpc}$, 7 billion particles) to give the best number statistics. 
We select galaxies within the redshift range $1.5<z<4$. For the version of the EAGLE simulation we use, the particle mass resolution for gas is $1.81\times10^{6}\,\rm{M_{\odot}}$, and the particle mass resolution for dark matter is $9.71\times10^{6}\,\rm{M_{\odot}}$. The resolution was chosen to marginally resolve the Jeans mass of the gas at the star formation threshold. The comoving (Plummer equivalent) gravitational softening length is $2.66\,\rm{kpc}$. Since EAGLE is a large box cosmological simulation, its resolution is low compared to the HR and SR FIRE zoom-ins.
\subsection{Modelled sub-millimeter flux densities for EAGLE galaxies}\label{sec:eagle_rt}
\begin{figure*} 
	\hspace{-0.5cm}
	\includegraphics[width=\columnwidth]{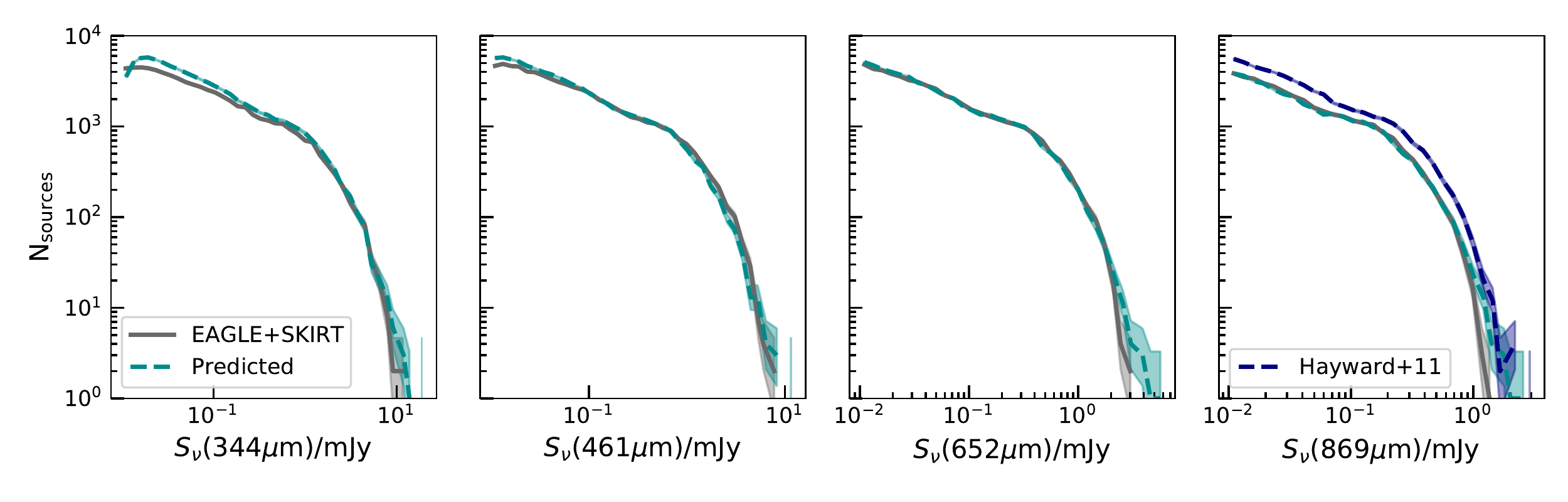}
	\hspace{-0.37cm}
	\caption{Number counts at $1.5<z<4$ for four sub-mm wavelengths, as predicted by the scaling relations from the following properties of the EAGLE galaxies: dust mass, stellar mass, SFR and redshift (dark cyan). In light cyan, we show Poisson uncertainties (calculated using single-sided errors drawn from \protect\cite{Gehrels1986} for cases where $N<100$). In dark grey, we show the number counts predicted by running the {\sc skirt} radiative transfer code on the EAGLE galaxies. The grey shaded region shows Poisson uncertainties. The number counts generated using our scaling relations are generally very similar to those yielded by performing radiative transfer on the EAGLE galaxies. This shows that our scaling relations can be applied to other simulations and semi-analytical or semi-empirical models and quickly generate robust predictions for number counts, even when these other simulations have different input physics and particle resolutions. We overlay the $870\,\mu\rm{m}$ number counts predicted using the scaling relation derived by \protect\cite{Hayward2011} in blue. These over-predict the EAGLE+{\sc skirt}-derived counts by $\sim20$ per cent.}
    \label{fig:alma_lf}
\end{figure*}
We draw synthetic fluxes for the EAGLE galaxies from the `DustyFluxes' table, which is part of the public database. These observed-frame fluxes were generated following the methods described in \cite{Camps2017}, which we summarise here. 
For each galaxy, gas and star particles were extracted within $30$ proper $\rm{kpc}$ of the centre of mass of the star particles. Although the EAGLE project also uses the {\sc{skirt}} radiative transfer code to generate galaxy SEDs, their methods of generating the intrinsic stellar emission differ from ours (primarily due to the lower particle mass resolution of EAGLE). Star particles with ages $<100\,\rm{Myr}$ and gas particles with non-zero SFRs were re-sampled into a number of sub-particles. These sub-particles have lower masses, drawn randomly from a mass distribution function chosen based on observations of molecular clouds in the Milky Way. Each sub-particle has a random formation time, assuming constant SFR over the previous $100\,\rm{Myr}$. Of the generated set of sub-particles, those with assigned ages $<10\,\rm{Myr}$ define a set of `star-forming' particles input separately to the radiative transfer. Older sub-particles were added to the set of star particles with ages $>100\,\rm{Myr}$. Gas sub-particles that have not yet formed stars were added to the set of gas particles with zero SFR. There were then three inputs to the {\sc{skirt}} radiative transfer: stars with ages $>90\,\rm{Myr}$, stars with ages $<10\,\rm{Myr}$, and gas.\\
\indent Within {\sc{skirt}}, dust was modelled based on the `cold' gas particles (those with a non-zero SFR or temperature $T<8000\,\rm{K}$), assuming a dust-to-metals mass fraction of $0.3$ and a \cite{Zubko2004} dust model. SEDs were assigned to stellar particles with ages $>90\,\rm{Myr}$ using a \cite{Charlot2003} models, taking into account mass, metallicity and age. Younger star particles were assigned starburst SEDs from the MAPPINGS III family \citep{Groves2008,Jonsson2010}, which aims to model the HII regions as well as the photodissociation regions (PDRs). Each young particle was also assigned a position based on the position and smoothing length of its parent particle. Since dust is partially modelled by the MAPPINGS implementation of the PDR, the implicit PDR dust masses were subtracted from the diffuse dust distribution surrounding the star-forming region (otherwise this dust would be `double counted'). The radiative transfer was then performed on an octree grid.

\subsection{Rapid predictions of number counts using the trained scaling relations}
In this section, we apply the scaling relations derived using FIRE galaxies (as described in Section \ref{sec:four_flux_densities}) to EAGLE galaxies, generating predicted flux densities in four ALMA bands for each EAGLE galaxy. We generate number counts using these predicted flux densities, as well as using the flux densities derived directly from the application of {\sc{skirt}} to EAGLE described in Section \ref{sec:eagle_rt}, and compare the resulting number counts.\\
\indent We first select EAGLE galaxies in the redshift range $1.5<z<4$. We input the relevant features for each of the $\sim80,000$ EAGLE snapshots in the sample to the scaling relations derived in Section \ref{sec:four_flux_densities}. Note that although EAGLE have used a different dust-to-metals mass ratio in their calculation of dust mass ($0.3$, compared to our $0.4$), this does not affect the robustness of our flux density estimates because the dust mass rather than the gas mass is used in our scaling relation. The process of predicting the flux densities that would be observed at the wavelengths of the various ALMA bands, for all EAGLE snapshots, from the dust mass, star formation rate, stellar mass and redshift takes just a few minutes on a laptop. \\
\indent We compare the flux densities predicted using our scaling relations to those predicted using {\sc{skirt}}. We show the median difference between model and {\sc{skirt}}-predicted flux density values, as a function of model-predicted flux density, in the middle panel of Figure \ref{fig:all_bands_calibration}. Encouragingly, there is no significant bias (i.e. the median offset is $\sim0$), except at the shortest wavelength studied, where {\sc{skirt}} predicts lower flux densities than our relations for the faintest sources. The scatter in this relation is shown in the lower panel of Figure \ref{fig:all_bands_calibration}. The scatter for EAGLE galaxies is low, within $0.2\,\rm{dex}$ across over two orders of magnitude of flux density, for all wavelengths studied. The success of our FIRE-derived scaling relations in predicting the flux densities for EAGLE galaxies is notable. It means that, for galaxies drawn from EAGLE and FIRE with a given set of physical parameters (dust mass, stellar mass, star formation rate and redshift), {\sc{skirt}} predicts very similar sub-millimeter flux density values, despite the different input physics, subgrid models and resolutions of the two simulations. This is encouraging for the broader application of the scaling relations. \\
\indent We display the number counts predicted by the scaling relations and those derived directly from EAGLE+{\sc skirt} in Figure \ref{fig:alma_lf}. These show excellent agreement, in general to within Poisson uncertainties (unlike those derived using the scaling relation derived by \citealt{Hayward2011}, which over-predict counts by $\sim20\%$ at all flux densities). We marginally over-predict bright sources, possibly due to \cite{Eddington1913} bias. The overall success of our scaling relations is particularly encouraging given that they were trained on simulations with different input physics and resolutions, and with slightly different radiative transfer setup.
This shows that our scaling relations can be applied to other simulations and semi-analytical or semi-empirical models to generate robust and fast predictions for number counts.

\section{Discussion}\label{sec:discussion}
Our FIRE-trained scaling relations are remarkably accurate in predicting sub-millimeter flux densities of EAGLE galaxies from global galaxy properties alone. They provide a promising path for rapid generation of sub-millimeter number counts where radiative transfer is infeasible. Here, we will discuss some limitations of our method and avenues for future work.\\
\indent Although the FIRE-2 simulations used here succeed in reproducing many observed galaxy properties, including the compact sizes of massive sub-millimeter galaxies \citep{Cochrane2019}, there remain some limitations to the model. Firstly, as shown in \cite{Parsotan2021}, the optical sizes of massive FIRE-2 galaxies are over-compact compared to the observationally-derived size-stellar mass relation, particularly at $z<2$. Less massive FIRE-2 galaxies do not display this issue \citep{El-Badry2016}. This is at least partially attributable to the lack of AGN feedback in the models; numerous works have shown that AGN-driven winds leave an impact on galaxy stellar sizes and densities \citep[e.g.][]{Choi2018,Zoldan2019,VanderVlugt2020}. Efforts to implement AGN feedback into the simulations have already begun \citep{Wellons2022,Angles-Alcazar}, and future work will explore the impact of AGN-driven winds on physical and observable galaxy sizes \citep{Cochrane2022c}. An AGN feedback implementation would impact the evolution of stellar mass and star formation rate in a simulated massive galaxy. However, in this work we derive scaling relations to predict sub-millimeter flux density given some input physical quantities, so this work should be robust to changes in the details of the feedback models. The success of our model in predicting sub-millimeter number counts of galaxies drawn from the EAGLE simulations, which do include thermal AGN feedback \citep{Schaye2015}, supports this. \\
\indent Secondly, although our radiative transfer method is well-established and sophisticated compared to simple dust screen models, we do not model dust production, growth or destruction in a self-consistent manner. In this paper we use a fixed dust-to-metals ratio of $0.4$; this assumption and slight variations on it are common in the post-processing of zoom and cosmological scale simulations \citep[e.g.][]{Camps2016,Liang2018,Liang2019a,Trayford2019,Cochrane2019,Parsotan2021,Popping2021}, though note that efforts to implement dust evolution in FIRE are ongoing \citep{Choban2022}. In S{\sc imba}, which does include a model for on-the-fly dust production and destruction (following \citealt{McKinnon2017}), average dust masses are a factor of $\sim2.5$ higher than if this fixed dust-to-metals ratio had been applied \citep{Lovell2020}. These leads to more bright sub-millimeter galaxies in S{\sc imba}. Nevertheless, since dust mass is an input parameter to our scaling relations, the derived relations should not be highly sensitive to the exact values. Of more concern is the spatially-invariant dust-to-metals ratio; observational results disagree about the validity of this assumption (see \citealt{Chiang2018} and references therein). If the geometry of the dust is important for the flux densities and unrealistic, our scaling relations could be biased. However, the success of our model in bypassing radiative transfer for EAGLE galaxies, which are simulated at lower spatial resolution, points towards this being of negligible importance at the wavelengths studied. \\
\indent We have limited the scope of this work to a small set of observed-frame sub-millimeter flux densities. As argued, sub-millimeter number counts and redshift distributions can provide an important constraint to galaxy formation models. Ideally, though, our model would be generalizable to shorter wavelengths and able to make fast predictions for full galaxy SEDs. In future work, we will explore machine learning-based approaches to this problem.

\section{Conclusions}\label{sec:conclusions}
In this paper, we have trained simple scaling relations between observable sub-millimeter flux densities of simulated high redshift galaxies and combinations of their key intrinsic physical properties, including star formation rate, dust mass, stellar mass and redshift. We use a set of high-resolution cosmological zoom-in simulations of massive, star-forming galaxies from the FIRE suite. We use {\sc skirt} to perform radiative transfer on the simulated galaxies, thereby incorporating a realistic star-dust geometry (assuming that the dust-to-metals mass ratio is constant and does not vary spatially). We generate observed-frame galaxy SEDs, from which we extract predictions for the emission that would be measured by ALMA bands $10$, $9$, $8$, and $7$ ($352-870\,\mu\rm{m}$). \\
\indent The primary aim of this project is to make use of our very high resolution and computationally expensive simulations to derive relations that can be applied to other models. Applications include large box simulations, for which running radiative transfer can be infeasible; lower-resolution simulations for which the dust-star geometry is inadequately resolved for robust SED prediction; and semi-analytic and semi-empirical models for which detailed detailed spatial distributions of stars and dust do not exist. The relations we derive between physical properties (dust mass, stellar mass, $\rm{SFR_{10}}$ and redshift) and $352-870\,\mu\rm{m}$ flux density have strong physical bases. At longer wavelengths, the dependence of flux density on dust mass increases, as would be expected as we move closer to the Rayleigh-Jeans regime. Star formation rate plays an increased role at shorter wavelengths, further from the Rayleigh-Jeans regime, where flux density is expected to depend super-linearly on dust temperature. At all wavelengths studied, flux density is only weakly dependent on stellar mass. The relations we derive are broadly consistent with previous work using both single halo idealised zooms and cosmological scale simulations; this indicates that global galaxy properties determine most of the variation in the sub-mm flux densities of galaxies.\\
\indent We apply our derived scaling relations to galaxies from EAGLE, a large box cosmological hydrodynamical simulation. Importantly, radiative transfer has already been performed on these EAGLE galaxies, with {\sc skirt} outputs provided as part of the publicly-available online database. We find that EAGLE sub-millimeter number counts derived using our simple scaling relations agree remarkably well with those constructed using the radiative transfer-derived flux densities. This provides confidence in the validity of the application of the scaling relations to simulations with very different resolutions and physics to FIRE-2. As highlighted by \cite{Cochrane2019}, UV-optical emission is expected to be more strongly dependent on the resolved ISM geometry and viewing angle; predictions for emission at these shorter wavelengths thus strongly motivate detailed radiative transfer. In future work, we will explore methods to predict the entire UV-FIR galaxy SED from intrinsic galaxy properties. 

\section*{Acknowledgements}
\indent We thank the anonymous reviewer, Caleb Choban and Desika Narayanan for helpful comments on earlier versions of this paper. The Flatiron Institute is supported by the Simons Foundation. RKC is grateful for the support of the John Harvard Distinguished Science Fellowship. DAA was supported in part by NSF grants AST-2009687 and AST-2108944, and CXO grant TM2-23006X. The FIRE project is supported by the NSF and TACC, by grants AST21010 and AST20016.

\section*{Data availability}
The data underlying this article will be shared on reasonable request to the corresponding author.

\bibliographystyle{mnras}
\bibliography{Edinburgh}

\begin{thebibliography}{}
\makeatletter
\relax
\def\mn@urlcharsother{\let\do\@makeother \do\$\do\&\do\#\do\^\do\_\do\%\do\~}
\def\mn@doi{\begingroup\mn@urlcharsother \@ifnextchar [ {\mn@doi@}
  {\mn@doi@[]}}
\def\mn@doi@[#1]#2{\def\@tempa{#1}\ifx\@tempa\@empty \href
  {http://dx.doi.org/#2} {doi:#2}\else \href {http://dx.doi.org/#2} {#1}\fi
  \endgroup}
\def\mn@eprint#1#2{\mn@eprint@#1:#2::\@nil}
\def\mn@eprint@arXiv#1{\href {http://arxiv.org/abs/#1} {{\tt arXiv:#1}}}
\def\mn@eprint@dblp#1{\href {http://dblp.uni-trier.de/rec/bibtex/#1.xml}
  {dblp:#1}}
\def\mn@eprint@#1:#2:#3:#4\@nil{\def\@tempa {#1}\def\@tempb {#2}\def\@tempc
  {#3}\ifx \@tempc \@empty \let \@tempc \@tempb \let \@tempb \@tempa \fi \ifx
  \@tempb \@empty \def\@tempb {arXiv}\fi \@ifundefined
  {mn@eprint@\@tempb}{\@tempb:\@tempc}{\expandafter \expandafter \csname
  mn@eprint@\@tempb\endcsname \expandafter{\@tempc}}}

\bibitem[\protect\citeauthoryear{Angl{\'{e}}s-Alc{\'{a}}zar, Dav{\'{e}},
  Faucher-Gigu{\`{e}}re, {\"{O}}zel  \& Hopkins}{Angl{\'{e}}s-Alc{\'{a}}zar
  et~al.}{2017a}]{Angl2018}
Angl{\'{e}}s-Alc{\'{a}}zar D.,  Dav{\'{e}} R.,  Faucher-Gigu{\`{e}}re C.-A.,
  {\"{O}}zel F.,   Hopkins P.~F.,  2017a, \mn@doi [MNRAS]
  {10.1093/mnras/stw2565}, 464, 2840

\bibitem[\protect\citeauthoryear{Angl{\'{e}}s-Alc{\'{a}}zar,
  Faucher-Gigu{\`{e}}re, Kere{\v{s}}, Hopkins, Quataert  \&
  Murray}{Angl{\'{e}}s-Alc{\'{a}}zar et~al.}{2017b}]{Angles-Alcazar2016}
Angl{\'{e}}s-Alc{\'{a}}zar D.,  Faucher-Gigu{\`{e}}re C.-A.,  Kere{\v{s}} D.,
  Hopkins P.~F.,  Quataert E.,   Murray N.,  2017b, \mn@doi [MNRAS]
  {10.1093/mnras/stx1517}, 470, 4698

\bibitem[\protect\citeauthoryear{Angl{\'{e}}s-Alc{\'{a}}zar,
  Faucher-Gigu{\`{e}}re, Quataert, Hopkins, Feldmann, Torrey, Wetzel  \&
  Kere{\v{s}}}{Angl{\'{e}}s-Alc{\'{a}}zar et~al.}{2017c}]{Angles-Alcazar2017}
Angl{\'{e}}s-Alc{\'{a}}zar D.,  Faucher-Gigu{\`{e}}re C.~A.,  Quataert E.,
  Hopkins P.~F.,  Feldmann R.,  Torrey P.,  Wetzel A.,   Kere{\v{s}} D.,
  2017c, \mn@doi [MNRAS] {10.1093/mnrasl/slx161}, 472, L109

\bibitem[\protect\citeauthoryear{Angl{\'{e}}s-Alc{\'{a}}zar, Mercedes-Feliz, Al
   \& Al}{Angl{\'{e}}s-Alc{\'{a}}zar et~al.}{2022}]{Angles-Alcazar}
Angl{\'{e}}s-Alc{\'{a}}zar D.,  Mercedes-Feliz J.,  Al E.,   Al E.,  2022, in
  prep.

\bibitem[\protect\citeauthoryear{Baes, Verstappen, {De Looze}, Fritz, Saftly,
  {Vidal P{\'{e}}rez}, Stalevski  \& Valcke}{Baes et~al.}{2011}]{Baes2011}
Baes M.,  Verstappen J.,  {De Looze} I.,  Fritz J.,  Saftly W.,  {Vidal
  P{\'{e}}rez} E.,  Stalevski M.,   Valcke S.,  2011, \mn@doi [ApJS]
  {10.1088/0067-0049/196/2/22}, 196

\bibitem[\protect\citeauthoryear{Barger, Cowie, Sanders, Fulton, Taniguchi,
  Sato, Kawara  \& Okuda}{Barger et~al.}{1998}]{Barger1998}
Barger A.~J.,  Cowie L.,  Sanders D.~B.,  Fulton E.,  Taniguchi Y.,  Sato Y.,
  Kawara K.,   Okuda H.,  1998, \mn@doi [Nature]
  {10.1016/j.physrep.2014.02.009}, 394, 248

\bibitem[\protect\citeauthoryear{Baugh, Lacey, Frenk, Granato, Silva, Bressan,
  Benson  \& Cole}{Baugh et~al.}{2005}]{Baugh2005}
Baugh C.~M.,  Lacey C.~G.,  Frenk C.~S.,  Granato G.~L.,  Silva L.,  Bressan
  A.,  Benson A.~J.,   Cole S.,  2005, \mn@doi [MNRAS]
  {10.1111/j.1365-2966.2004.08553.x}, 356, 1191

\bibitem[\protect\citeauthoryear{Blain, Smail, Ivison, Kneib  \& Frayer}{Blain
  et~al.}{2002}]{Blain2002}
Blain A.~W.,  Smail I.,  Ivison R.~J.,  Kneib J.~P.,   Frayer D.~T.,  2002,
  \mn@doi [Phys. Rep.] {10.1016/S0370-1573(02)00134-5}, 369, 111

\bibitem[\protect\citeauthoryear{Bouwens, Stefanon, Oesch, Illingworth,
  Nanayakkara, Roberts-Borsani, Labb{\'{e}}  \& Smit}{Bouwens
  et~al.}{2019}]{Bouwens2019}
Bouwens R.~J.,  Stefanon M.,  Oesch P.~A.,  Illingworth G.~D.,  Nanayakkara T.,
   Roberts-Borsani G.,  Labb{\'{e}} I.,   Smit R.,  2019, \mn@doi [ApJ]
  {10.3847/1538-4357/ab24c5}, 880, 25

\bibitem[\protect\citeauthoryear{Bouwens et~al.,}{Bouwens
  et~al.}{2020}]{Bouwens2020a}
Bouwens R.,  et~al., 2020, \mn@doi [ApJ] {10.3847/1538-4357/abb830}, 902, 112

\bibitem[\protect\citeauthoryear{Bouwens et~al.,}{Bouwens
  et~al.}{2021}]{Bouwens2021a}
Bouwens R.~J.,  et~al., 2021, \mn@doi [ApJ] {10.3847/1538-3881/abf83e}, 162, 47

\bibitem[\protect\citeauthoryear{Bouwens, Illingworth, Ellis, Oesch,
  Paulino-Afonso, Ribeiro  \& Stefanon}{Bouwens et~al.}{2022}]{Bouwens2022}
Bouwens R.~J.,  Illingworth G.,  Ellis R.~S.,  Oesch P.,  Paulino-Afonso A.,
  Ribeiro B.,   Stefanon M.,  2022, ApJ, 931, 81

\bibitem[\protect\citeauthoryear{Bruzual \& Charlot}{Bruzual \&
  Charlot}{2003}]{Charlot2003}
Bruzual G.,  Charlot S.,  2003, MNRAS, 344, 1000

\bibitem[\protect\citeauthoryear{{Calistro Rivera} et~al.,}{{Calistro Rivera}
  et~al.}{2018}]{Rivera2018}
{Calistro Rivera} G.,  et~al., 2018, \mn@doi [ApJ] {10.3847/1538-4357/aacffa},
  863, 56

\bibitem[\protect\citeauthoryear{Camps \& Baes}{Camps \&
  Baes}{2015}]{Camps2014}
Camps P.,  Baes M.,  2015, \mn@doi [Astronomy and Computing]
  {10.1016/j.ascom.2014.10.004}, 9, 20

\bibitem[\protect\citeauthoryear{Camps, Trayford, Baes, Theuns, Schaller  \&
  Schaye}{Camps et~al.}{2016}]{Camps2016}
Camps P.,  Trayford J.~W.,  Baes M.,  Theuns T.,  Schaller M.,   Schaye J.,
  2016, \mn@doi [MNRAS] {10.1093/mnras/stw1735}, 462, 1057

\bibitem[\protect\citeauthoryear{Camps et~al.,}{Camps et~al.}{2018}]{Camps2017}
Camps P.,  et~al., 2018, \mn@doi [ApJ] {10.3847/1538-4365/aaa24c}, 234, 20

\bibitem[\protect\citeauthoryear{Casey et~al.,}{Casey
  et~al.}{2018}]{Casey2018c}
Casey C.~M.,  et~al., 2018, \mn@doi [ApJ] {10.3847/1538-4357/aac82d}, 862, 77

\bibitem[\protect\citeauthoryear{Chabrier}{Chabrier}{2003}]{Chabrier2003}
Chabrier G.,  2003, PASP, 115, 763

\bibitem[\protect\citeauthoryear{Chen et~al.,}{Chen et~al.}{2017}]{Chen2017}
Chen C.-C.,  et~al., 2017, \mn@doi [ApJ] {10.3847/1538-4357/aa863a}, 846, 108

\bibitem[\protect\citeauthoryear{Chen et~al.,}{Chen et~al.}{2020}]{Chen2020}
Chen C.-C.,  et~al., 2020, \mn@doi [A\&A] {10.1051/0004-6361/201936286}, 635,
  A119

\bibitem[\protect\citeauthoryear{Chiang, Sandstrom, Chastenet, Johnson, Leroy
  \& Utomo}{Chiang et~al.}{2018}]{Chiang2018}
Chiang I.-D.,  Sandstrom K.~M.,  Chastenet J.,  Johnson L.~C.,  Leroy A.~K.,
  Utomo D.,  2018, \mn@doi [ApJ] {10.3847/1538-4357/aadc5f}, 865, 117

\bibitem[\protect\citeauthoryear{Choban, Hopkins, Sandstrom, Hayward  \&
  Faucher-Gigu{\`{e}}re}{Choban et~al.}{2022}]{Choban2022}
Choban C.~R.,  Hopkins D. K. P.~F.,  Sandstrom K.~M.,  Hayward C.~C.,
  Faucher-Gigu{\`{e}}re C.-A.,  2022, \mn@doi [MNRAS] {10.1093/mnras/stac1542},
  22, 1

\bibitem[\protect\citeauthoryear{Choi, Somerville, Ostriker, Naab  \&
  Hirschmann}{Choi et~al.}{2018}]{Choi2018}
Choi E.,  Somerville R.~S.,  Ostriker J.~P.,  Naab T.,   Hirschmann M.,  2018,
  \mn@doi [ApJ] {10.3847/1538-4357/aae076}, 866, 91

\bibitem[\protect\citeauthoryear{Cochrane \& Angles-Alcazar}{Cochrane \&
  Angles-Alcazar}{2022}]{Cochrane2022c}
Cochrane R.,  Angles-Alcazar D.,  2022, in prep.

\bibitem[\protect\citeauthoryear{Cochrane et~al.,}{Cochrane
  et~al.}{2019}]{Cochrane2019}
Cochrane R.~K.,  et~al., 2019, \mn@doi [MNRAS] {10.1093/mnras/stz1736}, 488,
  1779

\bibitem[\protect\citeauthoryear{Cochrane, Best, Smail, Ibar, Cheng, Swinbank,
  Molina  \& Sobral}{Cochrane et~al.}{2021}]{Cochrane2021}
Cochrane R.~K.,  Best P.~N.,  Smail I.,  Ibar E.,  Cheng C.,  Swinbank A.~M.,
  Molina J.,   Sobral D.,  2021, MNRAS, 503, 2622

\bibitem[\protect\citeauthoryear{Cochrane, Hayward  \& Angles-Alcazar}{Cochrane
  et~al.}{2022}]{Cochrane2022a}
Cochrane R.~K.,  Hayward C.~C.,   Angles-Alcazar D.,  2022, ApJL, 939, L27

\bibitem[\protect\citeauthoryear{Cole et~al.,}{Cole et~al.}{2000}]{Cole2000}
Cole S.,  et~al., 2000, \mn@doi [MNRAS] {10.1046/j.1365-8711.2001.04591.x},
  326, 255

\bibitem[\protect\citeauthoryear{Cowley, Lacey, Baugh, Cole, Frenk  \&
  Lagos}{Cowley et~al.}{2019}]{Cowley2018}
Cowley W.~I.,  Lacey C.~G.,  Baugh C.~M.,  Cole S.,  Frenk C.~S.,   Lagos C.
  d.~P.,  2019, MNRAS, 487, 3082

\bibitem[\protect\citeauthoryear{Crain et~al.,}{Crain et~al.}{2015}]{Crain2015}
Crain R.~A.,  et~al., 2015, \mn@doi [MNRAS] {10.1093/mnras/stv725}, 450, 1937

\bibitem[\protect\citeauthoryear{{Da Cunha} et~al.,}{{Da Cunha}
  et~al.}{2015}]{Cunha2015}
{Da Cunha} E.,  et~al., 2015, \mn@doi [ApJ] {10.1088/0004-637X/806/1/110}, 806,
  110

\bibitem[\protect\citeauthoryear{Dale}{Dale}{2015}]{Dale2015}
Dale J.~E.,  2015, \mn@doi [New Astronomy Reviews]
  {10.1016/j.newar.2015.06.001}, 68, 1

\bibitem[\protect\citeauthoryear{Dav{\'{e}}, Angl{\'{e}}s-Alc{\'{a}}zar,
  Narayanan, Li, Rafieferantsoa  \& Appleby}{Dav{\'{e}}
  et~al.}{2019}]{Dave2019}
Dav{\'{e}} R.,  Angl{\'{e}}s-Alc{\'{a}}zar D.,  Narayanan D.,  Li Q.,
  Rafieferantsoa M.~H.,   Appleby S.,  2019, \mn@doi [MNRAS]
  {10.1093/mnras/stz937}, 486, 2827

\bibitem[\protect\citeauthoryear{Draine \& Salpeter}{Draine \&
  Salpeter}{1979}]{Draine1979}
Draine B.~T.,  Salpeter E.~E.,  1979, \mn@doi [ApJ] {10.1086/157165}, 231, 77

\bibitem[\protect\citeauthoryear{Dudzeviciute et~al.,}{Dudzeviciute
  et~al.}{2020}]{Dudzeviciute2019}
Dudzeviciute U.,  et~al., 2020, MNRAS, 494, 3828

\bibitem[\protect\citeauthoryear{Dunlop et~al.,}{Dunlop
  et~al.}{2017}]{Dunlop2017}
Dunlop J.~S.,  et~al., 2017, \mn@doi [MNRAS] {10.1093/mnras/stw3088}, 466, 861

\bibitem[\protect\citeauthoryear{Dwek}{Dwek}{1998}]{Dwek1998}
Dwek E.,  1998, ApJ, 1, 643

\bibitem[\protect\citeauthoryear{Eddington}{Eddington}{1913}]{Eddington1913}
Eddington A.~S.,  1913, MNRAS, 73, 359

\bibitem[\protect\citeauthoryear{El-Badry, Wetzel, Geha, Hopkins, Kere{\v{s}},
  Chan  \& Faucher-Gigu{\`{e}}re}{El-Badry et~al.}{2016}]{El-Badry2016}
El-Badry K.,  Wetzel A.,  Geha M.,  Hopkins P.~F.,  Kere{\v{s}} D.,  Chan
  T.~K.,   Faucher-Gigu{\`{e}}re C.-A.,  2016, \mn@doi [ApJ]
  {10.3847/0004-637x/820/2/131}, 820, 131

\bibitem[\protect\citeauthoryear{Faucher-Gigu{\`{e}}re, Hopkins, Ker{\v{e}}s,
  Muratov, Quataert  \& Murray}{Faucher-Gigu{\`{e}}re
  et~al.}{2015}]{Faucher-Giguere2015}
Faucher-Gigu{\`{e}}re C.~A.,  Hopkins P.~F.,  Ker{\v{e}}s D.,  Muratov A.~L.,
  Quataert E.,   Murray N.,  2015, \mn@doi [MNRAS] {10.1093/mnras/stv336}, 449,
  987

\bibitem[\protect\citeauthoryear{Feldmann, Hopkins, Quataert,
  Faucher-Gigu{\`{e}}re  \& Ker{\v{e}}s}{Feldmann et~al.}{2016}]{Feldmann2016}
Feldmann R.,  Hopkins P.~F.,  Quataert E.,  Faucher-Gigu{\`{e}}re C.~A.,
  Ker{\v{e}}s D.,  2016, \mn@doi [MNRAS] {10.1093/mnrasl/slw014}, 458, L14

\bibitem[\protect\citeauthoryear{Feldmann, Quataert, Hopkins,
  Faucher-Gigu{\'{e}}re  \& Kere{\v{s}}}{Feldmann et~al.}{2017}]{Feldmann2017a}
Feldmann R.,  Quataert E.,  Hopkins P.~F.,  Faucher-Gigu{\'{e}}re C.~A.,
  Kere{\v{s}} D.,  2017, \mn@doi [MNRAS] {10.1093/mnras/stx1120}, 470, 1050

\bibitem[\protect\citeauthoryear{Finkelstein et~al.,}{Finkelstein
  et~al.}{2022}]{Finkelstein2022}
Finkelstein S.~L.,  et~al., 2022, \mn@doi [ApJ] {10.3847/1538-4357/ac3aed},
  928, 52

\bibitem[\protect\citeauthoryear{{Flores Vel{\'{a}}zquez} et~al.,}{{Flores
  Vel{\'{a}}zquez} et~al.}{2021}]{FloresVelazquez2021}
{Flores Vel{\'{a}}zquez} J.~A.,  et~al., 2021, \mn@doi [MNRAS]
  {10.1093/mnras/staa3893}, 501, 4812

\bibitem[\protect\citeauthoryear{Foreman-Mackey}{Foreman-Mackey}{2016}]{Foreman-Mackey2016}
Foreman-Mackey D.,  2016, JOSS, 24

\bibitem[\protect\citeauthoryear{Foreman-Mackey, Hogg, Lang  \&
  Goodman}{Foreman-Mackey et~al.}{2013}]{Foreman-Mackey2013}
Foreman-Mackey D.,  Hogg D.~W.,  Lang D.,   Goodman J.,  2013, \mn@doi [PASP]
  {10.1086/670067}, 125, 306

\bibitem[\protect\citeauthoryear{Fujimoto et~al.,}{Fujimoto
  et~al.}{2021}]{Fujimoto2021}
Fujimoto S.,  et~al., 2021, \mn@doi [ApJ] {10.3847/1538-4357/abd7ec}, 911, 99

\bibitem[\protect\citeauthoryear{Furlong}{Furlong}{2017}]{Furlong2015}
Furlong M.,  2017, MNRAS, 465, 722

\bibitem[\protect\citeauthoryear{Gaburov \& Nitadori}{Gaburov \&
  Nitadori}{2011}]{Gaburov2011}
Gaburov E.,  Nitadori K.,  2011, \mn@doi [MNRAS]
  {10.1111/j.1365-2966.2011.18313.x}, 414, 129

\bibitem[\protect\citeauthoryear{Garcia-Vergara, Hodge, Hennawi, Weiss,
  Wardlow, Myers  \& Hickox}{Garcia-Vergara et~al.}{2020}]{Garcia-Vergara2020}
Garcia-Vergara C.,  Hodge J.,  Hennawi J.~F.,  Weiss A.,  Wardlow J.,  Myers
  A.~D.,   Hickox R.,  2020, ApJ, 904

\bibitem[\protect\citeauthoryear{Gehrels}{Gehrels}{1986}]{Gehrels1986}
Gehrels N.,  1986, ApJ, 303, 336

\bibitem[\protect\citeauthoryear{Genel et~al.,}{Genel et~al.}{2014}]{Genel2014}
Genel S.,  et~al., 2014, \mn@doi [MNRAS] {10.1093/mnras/stu1654}, 445, 175

\bibitem[\protect\citeauthoryear{Granato, Lacey, Silva, Bressan, Baugh, Cole
  \& Frenk}{Granato et~al.}{2000}]{Granato2000}
Granato G.~L.,  Lacey C.~G.,  Silva L.,  Bressan A.,  Baugh C.~M.,  Cole S.,
  Frenk C.~S.,  2000, \mn@doi [ApJ] {10.1086/317032}, 542, 710

\bibitem[\protect\citeauthoryear{Groves, Dopita, Sutherland, Kewley, Fischera,
  Leitherer, Brandl  \& van Breugel}{Groves et~al.}{2008}]{Groves2008}
Groves B.,  Dopita M.~A.,  Sutherland R.~S.,  Kewley L.~J.,  Fischera J.,
  Leitherer C.,  Brandl B.,   van Breugel W.,  2008, \mn@doi [ApJS]
  {10.1086/528711}, 176, 438

\bibitem[\protect\citeauthoryear{Hafen et~al.,}{Hafen et~al.}{2019}]{Hafen2019}
Hafen Z.,  et~al., 2019, \mn@doi [MNRAS] {10.1093/mnras/stz1773}, 488, 1248

\bibitem[\protect\citeauthoryear{Hayward, Kere{\v{s}}, Jonsson, Narayanan, Cox
  \& Hernquist}{Hayward et~al.}{2011}]{Hayward2011}
Hayward C.~C.,  Kere{\v{s}} D.,  Jonsson P.,  Narayanan D.,  Cox T.~J.,
  Hernquist L.,  2011, \mn@doi [ApJ] {10.1088/0004-637X/743/2/159}, 743

\bibitem[\protect\citeauthoryear{Hayward, Jonsson, Kere{\v{s}}, Magnelli,
  Hernquist  \& Cox}{Hayward et~al.}{2012}]{Hayward2012}
Hayward C.~C.,  Jonsson P.,  Kere{\v{s}} D.,  Magnelli B.,  Hernquist L.,   Cox
  T.~J.,  2012, \mn@doi [MNRAS] {10.1111/j.1365-2966.2012.21254.x}, 424, 951

\bibitem[\protect\citeauthoryear{Hayward, Narayanan, Kere{\v{s}}, Jonsson,
  Hopkins, Cox  \& Hernquist}{Hayward et~al.}{2013a}]{Hayward2013}
Hayward C.~C.,  Narayanan D.,  Kere{\v{s}} D.,  Jonsson P.,  Hopkins P.~F.,
  Cox T.~J.,   Hernquist L.,  2013a, \mn@doi [MNRAS] {10.1093/mnras/sts222},
  428, 2529

\bibitem[\protect\citeauthoryear{Hayward, Behroozi, Somerville, Primack, Moreno
   \& Wechsler}{Hayward et~al.}{2013b}]{Hayward2013b}
Hayward C.~C.,  Behroozi P.~S.,  Somerville R.~S.,  Primack J.~R.,  Moreno J.,
   Wechsler R.~H.,  2013b, \mn@doi [MNRAS] {10.1093/mnras/stt1202}, 434, 2572

\bibitem[\protect\citeauthoryear{Hayward et~al.,}{Hayward
  et~al.}{2018}]{Hayward2018a}
Hayward C.~C.,  et~al., 2018, \mn@doi [MNRAS] {10.1093/mnras/sty304}, 476, 2278

\bibitem[\protect\citeauthoryear{Hayward et~al.,}{Hayward
  et~al.}{2021}]{Hayward2021}
Hayward C.~C.,  et~al., 2021, \mn@doi [MNRAS] {10.1093/mnras/stab246}, 502,
  2922

\bibitem[\protect\citeauthoryear{Hickox et~al.,}{Hickox
  et~al.}{2012}]{Hickox2012}
Hickox R.~C.,  et~al., 2012, \mn@doi [MNRAS]
  {10.1111/j.1365-2966.2011.20303.x}, 421, 284

\bibitem[\protect\citeauthoryear{Hodge \& da Cunha}{Hodge \&
  da~Cunha}{2020}]{Hodge2020}
Hodge J.~A.,  da Cunha E.,  2020, \mn@doi [Royal Society Open Science]
  {10.1098/rsos.200556}, 7, 200556

\bibitem[\protect\citeauthoryear{Hodge et~al.,}{Hodge et~al.}{2016}]{Hodge2016}
Hodge J.~A.,  et~al., 2016, \mn@doi [ApJ] {10.3847/1538-4357/833/1/103}, 833, 1

\bibitem[\protect\citeauthoryear{Hopkins}{Hopkins}{2015}]{Hopkins2015}
Hopkins P.~F.,  2015, \mn@doi [MNRAS] {10.1093/mnras/stv195}, 450, 53

\bibitem[\protect\citeauthoryear{Hopkins, Quataert  \& Murray}{Hopkins
  et~al.}{2011}]{Hopkins2011}
Hopkins P.~F.,  Quataert E.,   Murray N.,  2011, \mn@doi [MNRAS]
  {10.1111/j.1365-2966.2011.19306.x}, 417, 950

\bibitem[\protect\citeauthoryear{Hopkins, Narayanan  \& Murray}{Hopkins
  et~al.}{2013}]{Hopkins2013sf_criteria}
Hopkins P.~F.,  Narayanan D.,   Murray N.,  2013, \mn@doi [MNRAS]
  {10.1093/mnras/stt723}, 432, 2647

\bibitem[\protect\citeauthoryear{Hopkins, Kere{\v{s}}, O{\~{n}}orbe,
  Faucher-Gigu{\`{e}}re, Quataert, Murray  \& Bullock}{Hopkins
  et~al.}{2014}]{Hopkins2014}
Hopkins P.~F.,  Kere{\v{s}} D.,  O{\~{n}}orbe J.,  Faucher-Gigu{\`{e}}re C.~A.,
   Quataert E.,  Murray N.,   Bullock J.~S.,  2014, \mn@doi [MNRAS]
  {10.1093/mnras/stu1738}, 445, 581

\bibitem[\protect\citeauthoryear{Hopkins et~al.,}{Hopkins
  et~al.}{2018a}]{Hopkins2018feedback}
Hopkins P.~F.,  et~al., 2018a, \mn@doi [MNRAS] {10.1093/mnras/sty674}, 477,
  1578

\bibitem[\protect\citeauthoryear{Hopkins et~al.,}{Hopkins
  et~al.}{2018b}]{Hopkins2017}
Hopkins P.~F.,  et~al., 2018b, MNRAS, 480, 800

\bibitem[\protect\citeauthoryear{Hughes et~al.,}{Hughes
  et~al.}{1998}]{Hughes1998}
Hughes D.~H.,  et~al., 1998, \mn@doi [Nature] {10.1038/28328}, 394, 241

\bibitem[\protect\citeauthoryear{James, Dunne, Eales  \& Edmunds}{James
  et~al.}{2002}]{James2002}
James A.,  Dunne L.,  Eales S.,   Edmunds M.~G.,  2002, \mn@doi [MNRAS]
  {10.1046/j.1365-8711.2002.05660.x}, 335, 753

\bibitem[\protect\citeauthoryear{Jonsson, Groves  \& Cox}{Jonsson
  et~al.}{2010}]{Jonsson2010}
Jonsson P.,  Groves B.~A.,   Cox T.~J.,  2010, \mn@doi [MNRAS]
  {10.1111/j.1365-2966.2009.16087.x}, 403, 17

\bibitem[\protect\citeauthoryear{Kere{\v{s}}, Katz, Fardal, Dav{\'{e}}  \&
  Weinberg}{Kere{\v{s}} et~al.}{2009}]{Keres2009}
Kere{\v{s}} D.,  Katz N.,  Fardal M.,  Dav{\'{e}} R.,   Weinberg D.~H.,  2009,
  \mn@doi [MNRAS] {10.1111/j.1365-2966.2009.14541.x}, 395, 160

\bibitem[\protect\citeauthoryear{Kroupa}{Kroupa}{2002}]{Kroupa2002}
Kroupa P.,  2002, Science, 295, 82

\bibitem[\protect\citeauthoryear{Krumholz \& Gnedin}{Krumholz \&
  Gnedin}{2011}]{Krumholz2011}
Krumholz M.~R.,  Gnedin N.~Y.,  2011, \mn@doi [ApJ]
  {10.1088/0004-637X/729/1/36}, 729, 36

\bibitem[\protect\citeauthoryear{Lacey et~al.,}{Lacey et~al.}{2016}]{Lacey2016}
Lacey C.~G.,  et~al., 2016, \mn@doi [MNRAS] {10.1093/mnras/stw1888}, 462, 3854

\bibitem[\protect\citeauthoryear{Leitherer et~al.,}{Leitherer
  et~al.}{1999}]{Leitherer1999}
Leitherer C.,  et~al., 1999, ApJS, 123, 3

\bibitem[\protect\citeauthoryear{Liang, Feldmann, Faucher-Gigu{\`{e}}re,
  Kere{\v{s}}, Hopkins, Hayward, Quataert  \& Scoville}{Liang
  et~al.}{2018}]{Liang2018}
Liang L.,  Feldmann R.,  Faucher-Gigu{\`{e}}re C.-A.,  Kere{\v{s}} D.,  Hopkins
  P.~F.,  Hayward C.~C.,  Quataert E.,   Scoville N.~Z.,  2018, \mn@doi [MNRAS]
  {10.1093/mnrasl/sly071}, 88, 83

\bibitem[\protect\citeauthoryear{Liang et~al.,}{Liang
  et~al.}{2019}]{Liang2019a}
Liang L.,  et~al., 2019, \mn@doi [MNRAS] {10.1093/mnras/stz2134}, 489, 1397

\bibitem[\protect\citeauthoryear{Lovell, Geach, Dav{\'{e}}, Narayanan  \&
  Li}{Lovell et~al.}{2021}]{Lovell2020}
Lovell C.~C.,  Geach J.~E.,  Dav{\'{e}} R.,  Narayanan D.,   Li Q.,  2021,
  MNRAS, 502, 772

\bibitem[\protect\citeauthoryear{Ma, Hopkins, Faucher-Gigu{\`{e}}re, Zolman,
  Muratov, Kere{\v{s}}  \& Quataert}{Ma et~al.}{2016}]{Ma2016}
Ma X.,  Hopkins P.~F.,  Faucher-Gigu{\`{e}}re C.~A.,  Zolman N.,  Muratov
  A.~L.,  Kere{\v{s}} D.,   Quataert E.,  2016, \mn@doi [MNRAS]
  {10.1093/mnras/stv2659}, 456, 2140

\bibitem[\protect\citeauthoryear{Ma et~al.,}{Ma et~al.}{2018}]{Ma2017}
Ma X.,  et~al., 2018, MNRAS, 478, 1694

\bibitem[\protect\citeauthoryear{Magnelli et~al.,}{Magnelli
  et~al.}{2012}]{Magnelli2012}
Magnelli B.,  et~al., 2012, \mn@doi [A\&A] {10.1051/0004-6361/201118312}, 539,
  A155

\bibitem[\protect\citeauthoryear{Magnelli et~al.,}{Magnelli
  et~al.}{2020}]{Magnelli2020}
Magnelli B.,  et~al., 2020, \mn@doi [ApJ] {10.3847/1538-4357/ab7897}, 892, 66

\bibitem[\protect\citeauthoryear{Marrone et~al.,}{Marrone
  et~al.}{2018}]{Marrone2018}
Marrone D.~P.,  et~al., 2018, \mn@doi [Nature] {10.1038/nature24629}, 553, 51

\bibitem[\protect\citeauthoryear{McAlpine et~al.,}{McAlpine
  et~al.}{2016}]{McAlpine2015}
McAlpine S.,  et~al., 2016, \mn@doi [Astronomy and Computing]
  {10.1016/j.ascom.2016.02.004}, 15, 72

\bibitem[\protect\citeauthoryear{McAlpine et~al.,}{McAlpine
  et~al.}{2019}]{McAlpine2019}
McAlpine S.,  et~al., 2019, MNRAS, 488, 2440

\bibitem[\protect\citeauthoryear{McKinney, Hayward, Rosenthal,
  Martinez-Galarza, Pope, Sajina  \& Smith}{McKinney
  et~al.}{2021}]{McKinney2021}
McKinney J.,  Hayward C.~C.,  Rosenthal L.~J.,  Martinez-Galarza J.~R.,  Pope
  A.,  Sajina A.,   Smith H.~A.,  2021, ApJ, 921, 55

\bibitem[\protect\citeauthoryear{McKinnon, Torrey, Vogelsberger, Hayward  \&
  Marinacci}{McKinnon et~al.}{2017}]{McKinnon2017}
McKinnon R.,  Torrey P.,  Vogelsberger M.,  Hayward C.~C.,   Marinacci F.,
  2017, \mn@doi [MNRAS] {10.1093/mnras/stx467}, 468, 1505

\bibitem[\protect\citeauthoryear{Miettinen et~al.,}{Miettinen
  et~al.}{2017}]{Miettinen2017}
Miettinen O.,  et~al., 2017, \mn@doi [A\&A] {10.1051/0004-6361/201730762}, 606,
  1

\bibitem[\protect\citeauthoryear{Miller et~al.,}{Miller
  et~al.}{2018}]{Miller2018}
Miller T.~B.,  et~al., 2018, Nature, 556, 469

\bibitem[\protect\citeauthoryear{Muratov, Kere{\v{s}}, Faucher-Gigu{\`{e}}re,
  Hopkins, Quataert  \& Murray}{Muratov et~al.}{2015}]{Muratov2015a}
Muratov A.~L.,  Kere{\v{s}} D.,  Faucher-Gigu{\`{e}}re C.~A.,  Hopkins P.~F.,
  Quataert E.,   Murray N.,  2015, \mn@doi [MNRAS] {10.1093/mnras/stv2126},
  454, 2691

\bibitem[\protect\citeauthoryear{Narayanan et~al.,}{Narayanan
  et~al.}{2010}]{Narayanan2010}
Narayanan D.,  et~al., 2010, \mn@doi [MNRAS]
  {10.1111/j.1365-2966.2010.16997.x}, 407, 1701

\bibitem[\protect\citeauthoryear{Narayanan et~al.,}{Narayanan
  et~al.}{2015}]{Narayanan2015}
Narayanan D.,  et~al., 2015, \mn@doi [Nature] {10.1038/nature15383}, 525, 496

\bibitem[\protect\citeauthoryear{Narayanan et~al.,}{Narayanan
  et~al.}{2021}]{Narayanan2020}
Narayanan D.,  et~al., 2021, ApJS, 252, 12

\bibitem[\protect\citeauthoryear{Oesch, Bouwens, Illingworth, Labb{\'{e}}  \&
  Stefanon}{Oesch et~al.}{2018}]{Oesch2018}
Oesch P.~A.,  Bouwens R.~J.,  Illingworth G.~D.,  Labb{\'{e}} I.,   Stefanon
  M.,  2018, \mn@doi [ApJ] {10.3847/1538-4357/aab03f}, 855, 105

\bibitem[\protect\citeauthoryear{Pandya et~al.,}{Pandya
  et~al.}{2021}]{Pandya2021}
Pandya V.,  et~al., 2021, \mn@doi [MNRAS] {10.1093/mnras/stab2714}, 508, 2979

\bibitem[\protect\citeauthoryear{Parsotan, Cochrane, Hayward,
  Angl{\'{e}}s-Alc{\'{a}}zar, Feldmann, Faucher-Gigu{\`{e}}re, Wellons  \&
  Hopkins}{Parsotan et~al.}{2021}]{Parsotan2021}
Parsotan T.,  Cochrane R.~K.,  Hayward C.~C.,  Angl{\'{e}}s-Alc{\'{a}}zar D.,
  Feldmann R.,  Faucher-Gigu{\`{e}}re C.~A.,  Wellons S.,   Hopkins P.~F.,
  2021, \mn@doi [MNRAS] {10.1093/mnras/staa3765}, 501, 1591

\bibitem[\protect\citeauthoryear{Pillepich et~al.,}{Pillepich
  et~al.}{2018}]{Pillepich2018}
Pillepich A.,  et~al., 2018, \mn@doi [MNRAS] {10.1093/mnras/stx3112}, 475, 648

\bibitem[\protect\citeauthoryear{Popping et~al.,}{Popping
  et~al.}{2022}]{Popping2021}
Popping G.,  et~al., 2022, MNRAS, 510, 3321

\bibitem[\protect\citeauthoryear{Rowlands, Gomez, Dunne,
  Arag{\'{o}}n-Salamanca, Dye, Maddox, da Cunha  \& van~der Werf}{Rowlands
  et~al.}{2014}]{Rowlands2014}
Rowlands K.,  Gomez H.~L.,  Dunne L.,  Arag{\'{o}}n-Salamanca A.,  Dye S.,
  Maddox S.,  da Cunha E.,   van~der Werf P.,  2014, \mn@doi [MNRAS]
  {10.1093/mnras/stu605}, 441, 1040

\bibitem[\protect\citeauthoryear{Rupke}{Rupke}{2018}]{Rupke2018}
Rupke D. S.~N.,  2018, \mn@doi [Galaxies] {10.3390/galaxies6040138}, 6, 138

\bibitem[\protect\citeauthoryear{Rybak, McKean, Vegetti, Andreani  \&
  White}{Rybak et~al.}{2015a}]{Rybak2015a}
Rybak M.,  McKean J.~P.,  Vegetti S.,  Andreani P.,   White S.~D.,  2015a,
  \mn@doi [MNRAS] {10.1093/mnrasl/slv058}, 451, L40

\bibitem[\protect\citeauthoryear{Rybak, Vegetti, McKean, Andreani  \&
  White}{Rybak et~al.}{2015b}]{Rybak2015}
Rybak M.,  Vegetti S.,  McKean J.~P.,  Andreani P.,   White S.~D.,  2015b,
  \mn@doi [MNRAS] {10.1093/mnrasl/slv092}, 453, L26

\bibitem[\protect\citeauthoryear{Rybak, Hodge, Vegetti, van~der Werf, Andreani,
  Graziani  \& McKean}{Rybak et~al.}{2020}]{Rybak2020}
Rybak M.,  Hodge J.~A.,  Vegetti S.,  van~der Werf P.,  Andreani P.,  Graziani
  L.,   McKean J.~P.,  2020, \mn@doi [MNRAS] {10.1093/mnras/staa879}, 494, 5542

\bibitem[\protect\citeauthoryear{Safarzadeh, Hayward  \& Ferguson}{Safarzadeh
  et~al.}{2017}]{Safarzadeh2017}
Safarzadeh M.,  Hayward C.~C.,   Ferguson H.~C.,  2017, \mn@doi [ApJ]
  {10.3847/1538-4357/aa6c5b}, 840, 15

\bibitem[\protect\citeauthoryear{Schaller et~al.,}{Schaller
  et~al.}{2015}]{Schaller2015}
Schaller M.,  et~al., 2015, \mn@doi [MNRAS] {10.1093/mnras/stv1067}, 451, 1247

\bibitem[\protect\citeauthoryear{Schaye et~al.,}{Schaye
  et~al.}{2015}]{Schaye2015}
Schaye J.,  et~al., 2015, \mn@doi [MNRAS] {10.1093/mnras/stu2058}, 446, 521

\bibitem[\protect\citeauthoryear{Simpson et~al.,}{Simpson
  et~al.}{2014}]{Simpson2014}
Simpson J.~M.,  et~al., 2014, \mn@doi [ApJ] {10.1088/0004-637X/788/2/125}, 788

\bibitem[\protect\citeauthoryear{Smail, Ivison  \& Blain}{Smail
  et~al.}{1997}]{Smail1997}
Smail I.,  Ivison R.~J.,   Blain A.~W.,  1997, ApJ, 490, L5

\bibitem[\protect\citeauthoryear{Smail et~al.,}{Smail et~al.}{2021}]{Smail2020}
Smail I.,  et~al., 2021, MNRAS, 502, 3426

\bibitem[\protect\citeauthoryear{Somerville \& Dav{\'{e}}}{Somerville \&
  Dav{\'{e}}}{2015}]{Somerville}
Somerville R.~S.,  Dav{\'{e}} R.,  2015, \mn@doi [ARA\&A]
  {10.1146/annurev-astro-082812-140951}, 53, 51

\bibitem[\protect\citeauthoryear{Somerville, Gilmore, Primack  \&
  Dom{\'{i}}nguez}{Somerville et~al.}{2012}]{Somerville2012}
Somerville R.~S.,  Gilmore R.~C.,  Primack J.~R.,   Dom{\'{i}}nguez A.,  2012,
  \mn@doi [MNRAS] {10.1111/j.1365-2966.2012.20490.x}, 423, 1992

\bibitem[\protect\citeauthoryear{Sparre, Hayward, Feldmann,
  Faucher-Gigu{\`{e}}re, Muratov, Kere{\v{s}}  \& Hopkins}{Sparre
  et~al.}{2017}]{Sparre2017}
Sparre M.,  Hayward C.~C.,  Feldmann R.,  Faucher-Gigu{\`{e}}re C.~A.,  Muratov
  A.~L.,  Kere{\v{s}} D.,   Hopkins P.~F.,  2017, \mn@doi [MNRAS]
  {10.1093/mnras/stw3011}, 466, 88

\bibitem[\protect\citeauthoryear{Springel et~al.,}{Springel
  et~al.}{2005}]{Springel2005}
Springel V.,  et~al., 2005, \mn@doi [Nature] {10.1038/nature03597}, 435, 629

\bibitem[\protect\citeauthoryear{Stach et~al.,}{Stach et~al.}{2018}]{Stach2018}
Stach S.~M.,  et~al., 2018, \mn@doi [ApJ] {10.3847/1538-4357/aac5e5}, 860, 161

\bibitem[\protect\citeauthoryear{Stach et~al.,}{Stach
  et~al.}{2019}]{Stach2019b}
Stach S.~M.,  et~al., 2019, \mn@doi [MNRAS] {10.1093/mnras/stz1536}, 487, 4648

\bibitem[\protect\citeauthoryear{Stach et~al.,}{Stach et~al.}{2021}]{Stach2021}
Stach S.~M.,  et~al., 2021, MNRAS, 504, 172

\bibitem[\protect\citeauthoryear{Stefanon, Bouwens, Labb{\'{e}}, Illingworth,
  Gonzalez  \& Oesch}{Stefanon et~al.}{2021}]{Stefanon2021a}
Stefanon M.,  Bouwens R.~J.,  Labb{\'{e}} I.,  Illingworth G.~D.,  Gonzalez V.,
    Oesch P.~A.,  2021, \mn@doi [ApJ] {10.3847/1538-4357/ac1bb6}, 922, 29

\bibitem[\protect\citeauthoryear{Steinacker, Baes  \& Gordon}{Steinacker
  et~al.}{2013}]{Steinacker2013}
Steinacker J.,  Baes M.,   Gordon K.,  2013, \mn@doi [ARA\&A]
  {10.1146/annurev-astro-082812-141042}, pp 63--105

\bibitem[\protect\citeauthoryear{Swinbank et~al.,}{Swinbank
  et~al.}{2014}]{Swinbank2014}
Swinbank A.~M.,  et~al., 2014, \mn@doi [MNRAS] {10.1093/mnras/stt2273}, 438,
  1267

\bibitem[\protect\citeauthoryear{Tacchella et~al.,}{Tacchella
  et~al.}{2022}]{Tacchella2022}
Tacchella S.,  et~al., 2022, \mn@doi [ApJ] {10.3847/1538-4357/ac4cad}, 927, 170

\bibitem[\protect\citeauthoryear{Tadaki et~al.,}{Tadaki
  et~al.}{2017}]{Tadaki2017}
Tadaki K.-i.,  et~al., 2017, \mn@doi [ApJL] {10.3847/2041-8213/aa7338}, 841,
  L25

\bibitem[\protect\citeauthoryear{Tielens, McKee, Seab  \& Hollenbach}{Tielens
  et~al.}{1994}]{Tielens1994}
Tielens A. G. G.~M.,  McKee C.~F.,  Seab C.~G.,   Hollenbach D.~J.,  1994,
  \mn@doi [ApJ] {10.1086/174488}, 431, 321

\bibitem[\protect\citeauthoryear{Trayford \& Schaye}{Trayford \&
  Schaye}{2019}]{Trayford2019}
Trayford J.~W.,  Schaye J.,  2019, \mn@doi [MNRAS] {10.1093/mnras/stz757}, 485,
  5715

\bibitem[\protect\citeauthoryear{Trayford et~al.,}{Trayford
  et~al.}{2015}]{Trayford2015}
Trayford J.~W.,  et~al., 2015, \mn@doi [MNRAS] {10.1093/mnras/stv1461}, 452,
  2879

\bibitem[\protect\citeauthoryear{Trayford et~al.,}{Trayford
  et~al.}{2017}]{Trayford2017}
Trayford J.~W.,  et~al., 2017, MNRAS, 470, 771

\bibitem[\protect\citeauthoryear{{Van der Vlugt} et~al.,}{{Van der Vlugt}
  et~al.}{2021}]{VanderVlugt2020}
{Van der Vlugt} D.,  et~al., 2021, ApJ, 907, 5

\bibitem[\protect\citeauthoryear{Villaescusa-Navarro
  et~al.,}{Villaescusa-Navarro et~al.}{2021}]{Villaescusa-Navarro2021}
Villaescusa-Navarro F.,  et~al., 2021, \mn@doi [ApJ]
  {10.3847/1538-4357/abf7ba}, 915, 71

\bibitem[\protect\citeauthoryear{Villaescusa-Navarro
  et~al.,}{Villaescusa-Navarro et~al.}{2022}]{Villaescusa-Navarro2022b}
Villaescusa-Navarro F.,  et~al., 2022, arXiv:2201.01300

\bibitem[\protect\citeauthoryear{Vogelsberger et~al.,}{Vogelsberger
  et~al.}{2014}]{Vogelsberger2014a}
Vogelsberger M.,  et~al., 2014, \mn@doi [MNRAS] {10.1093/mnras/stu1536}, 444,
  1518

\bibitem[\protect\citeauthoryear{Wang, Pearson, Cowley, Trayford,
  B{\'{e}}thermin, Gruppioni, Hurley  \& Micha{\l}owski}{Wang
  et~al.}{2019}]{Wang2019c}
Wang L.,  Pearson W.~J.,  Cowley W.,  Trayford J.~W.,  B{\'{e}}thermin M.,
  Gruppioni C.,  Hurley P.,   Micha{\l}owski M.~J.,  2019, \mn@doi [A\&A]
  {10.1051/0004-6361/201834093}, 624, 1

\bibitem[\protect\citeauthoryear{Weinberger et~al.,}{Weinberger
  et~al.}{2017}]{Weinberger2017}
Weinberger R.,  et~al., 2017, \mn@doi [MNRAS] {10.1093/mnras/stw2944}, 465,
  3291

\bibitem[\protect\citeauthoryear{Weingartner \& Draine}{Weingartner \&
  Draine}{2001}]{Weingartner2001}
Weingartner J.~C.,  Draine B.~T.,  2001, \mn@doi [ApJ] {10.1086/318651}, 548,
  296

\bibitem[\protect\citeauthoryear{Wellons et~al.,}{Wellons
  et~al.}{2022}]{Wellons2022}
Wellons S.,  et~al., 2022, eprint arXiv:2203.06201

\bibitem[\protect\citeauthoryear{Wilkinson et~al.,}{Wilkinson
  et~al.}{2017}]{Wilkinson2016}
Wilkinson A.,  et~al., 2017, MNRAS, 464, 1380

\bibitem[\protect\citeauthoryear{Zavala et~al.,}{Zavala
  et~al.}{2021}]{Zavala2021}
Zavala J.~A.,  et~al., 2021, ApJ, 909, 165

\bibitem[\protect\citeauthoryear{Zoldan, {De Lucia}, Xie, Fontanot  \&
  Hirschmann}{Zoldan et~al.}{2019}]{Zoldan2019}
Zoldan A.,  {De Lucia} G.,  Xie L.,  Fontanot F.,   Hirschmann M.,  2019,
  \mn@doi [MNRAS] {10.1093/mnras/stz1670}, 487, 5649

\bibitem[\protect\citeauthoryear{Zubko, Dwek  \& Arendt}{Zubko
  et~al.}{2004}]{Zubko2004}
Zubko V.,  Dwek E.,   Arendt R.~G.,  2004, \mn@doi [ApJS] {10.1086/528711},
  152, 211

\bibitem[\protect\citeauthoryear{van~de Voort, Quataert, Hopkins, Kere{\v{s}}
  \& Faucher-Gigu{\'{e}}re}{van~de Voort et~al.}{2015}]{VandeVoort2015}
van~de Voort F.,  Quataert E.,  Hopkins P.~F.,  Kere{\v{s}} D.,
  Faucher-Gigu{\'{e}}re C.~A.,  2015, \mn@doi [MNRAS] {10.1093/mnras/stu2404},
  447, 140

\makeatother
\end{thebibliography}

\appendix
\section{Resolution tests}\label{sec:appendix_res_tests}
\begin{figure*} 
	\includegraphics[scale=0.58]{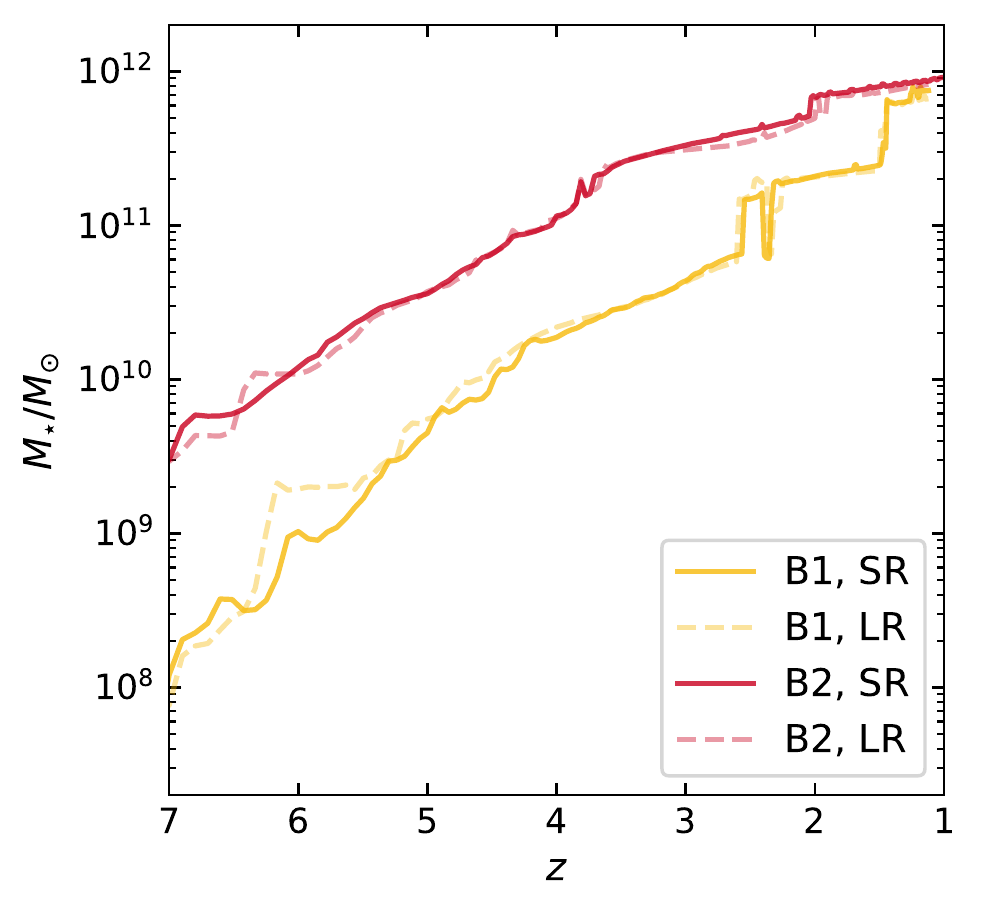}
	\includegraphics[scale=0.58]{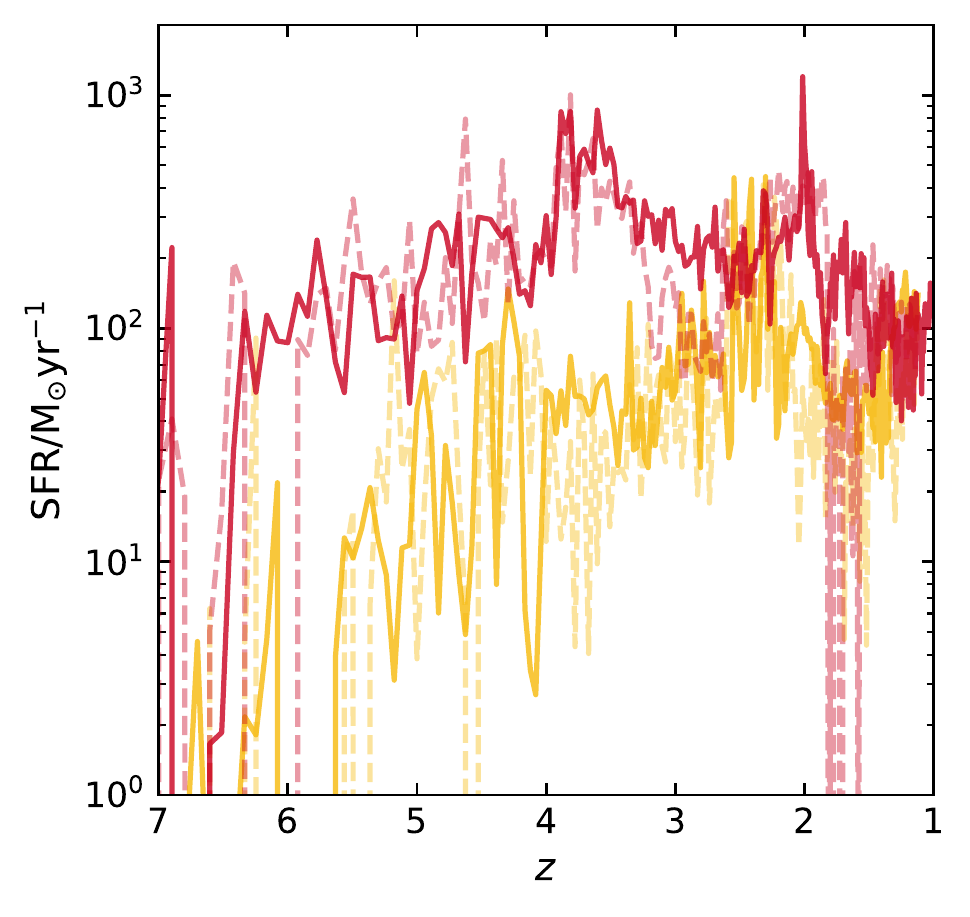}
	\includegraphics[scale=0.58]{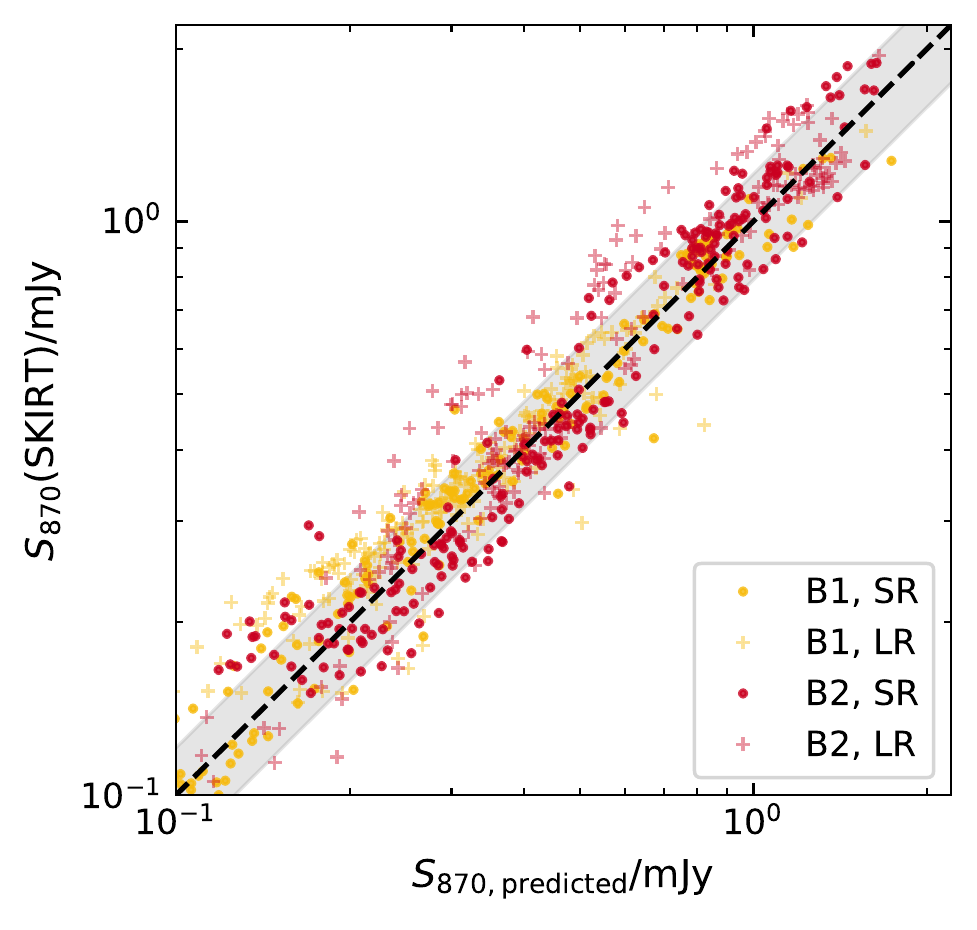}
	\caption{Convergence tests run on haloes B1 and B2. The solid lines show the evolution of stellar mass (left) and instantaneous star formation rate (center) for haloes B1 (yellow) and B2 (red), simulated at our standard resolution (SR). The dashed lines show the evolution of the same quantities for lower-resolution (LR) simulations of the same haloes. The simulations appear converged at the lower resolution, which is used for the most massive haloes in our sample, C1 and C2. The right-hand panel shows the {\sc{skirt}}-predicted observed-frame $870\,\mu\rm{m}$ flux densities versus our predicted values from the relation derived in Section \ref{sec:flux_predictions_3param} for both the standard- and low-resolution simulations of B1 and B2.}
    \label{fig:res_tests}
\end{figure*}
We perform convergence tests on the FIRE simulations using haloes B1 and B2. We re-run the zoom-in simulations at LR, to complement the SR simulations used in this work. We find excellent convergence of key properties, including stellar mass and star formation rate (see Figure \ref{fig:res_tests}). We also re-run the radiative transfer on the LR simulations and find no substantive differences in the relationships between sub-mm flux density and dust mass or SFR. In the right-hand panel of Figure \ref{fig:res_tests}, we plot {\sc{skirt}}-predicted observed-frame $870\,\mu\rm{m}$ flux densities for the SR and LR runs versus the values predicted from our fit to the standard resolution haloes, presented in Section \ref{sec:flux_predictions_3param}. The consistency between {\sc{skirt}} fluxes derived from the SR and LR runs indicates that the lowest resolution used in this paper (LR) is sufficient for our purposes.

\section{Convergence tests with radiative transfer}\label{sec:appendix_skirt_tests}
\begin{figure*} 
	\centering
        \begin{multicols}{2}
            \subcaptionbox{photon number\label{fig:1}}{\includegraphics[width=0.9\linewidth]{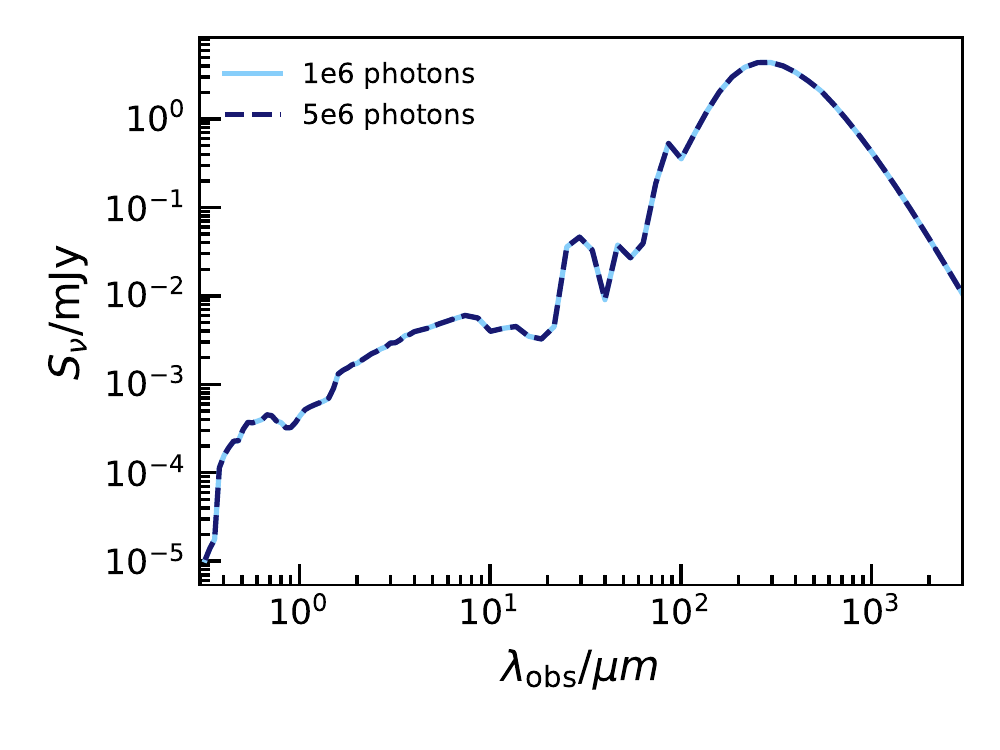}}\par 
            \hspace{-2cm}
            \subcaptionbox{maximum fraction of dust in one cell \label{fig:2}}{\includegraphics[width=0.9\linewidth]{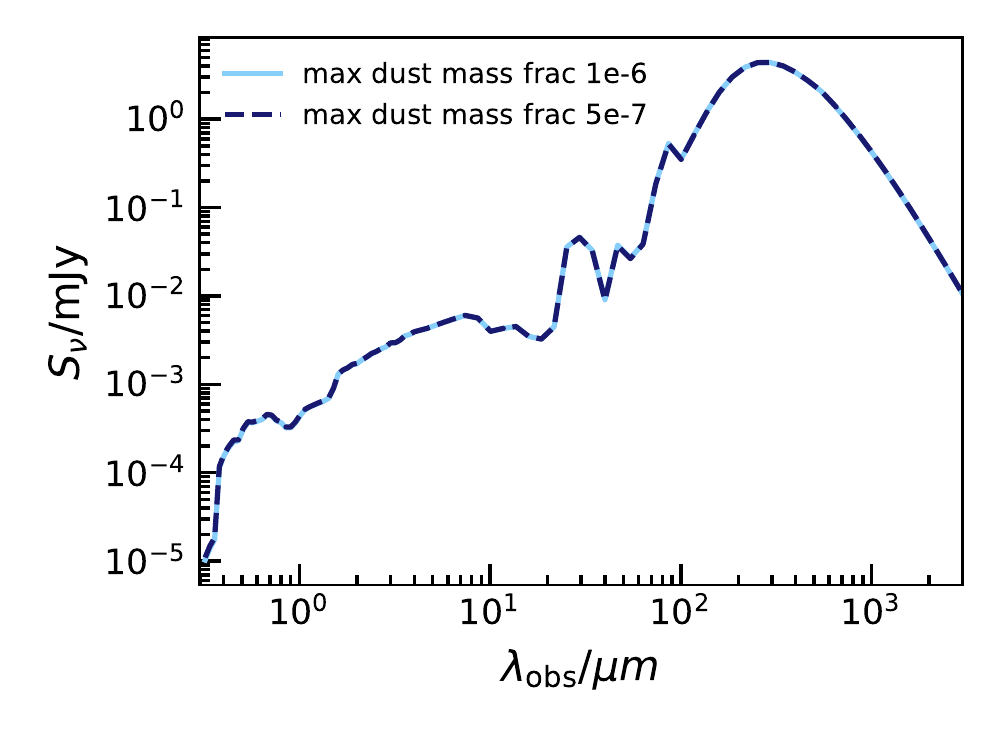}}\par
        \end{multicols}
        \begin{multicols}{2}
            \subcaptionbox{minimum dust grid level\label{fig:3}}{\includegraphics[width=0.9\linewidth]{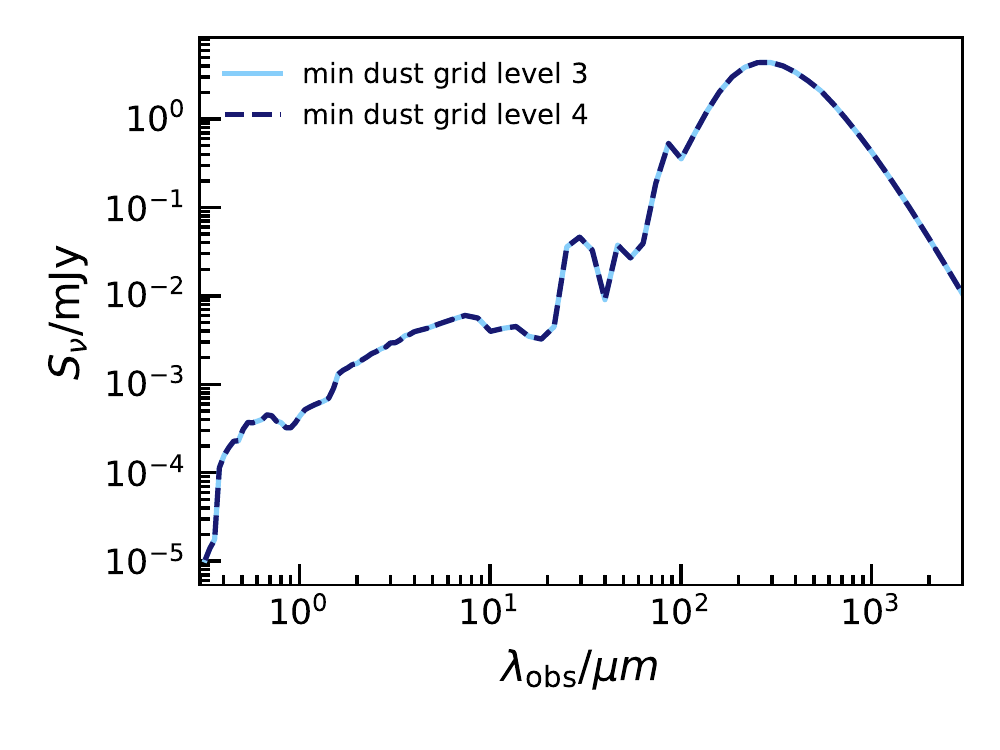}}\par
            \hspace{-2cm}
            \subcaptionbox{maximum dust grid level\label{fig:4}}{\includegraphics[width=0.9\linewidth]{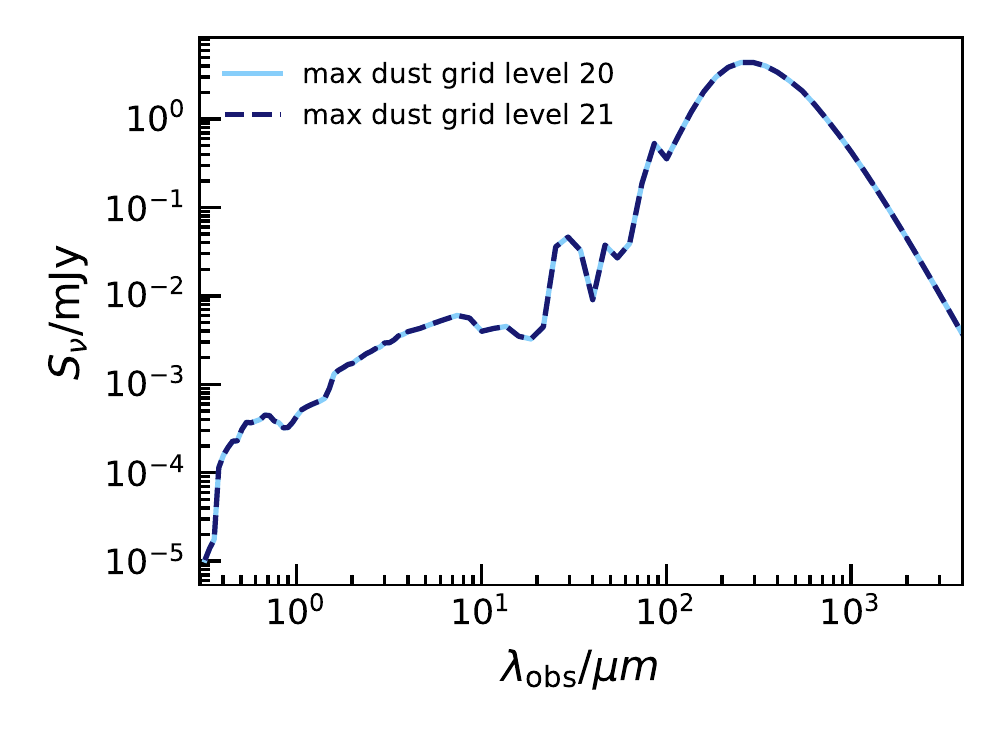}}\par 
        \end{multicols}
    \caption{Convergence tests performed on a single halo (A1 at $z=3$) with {\sc skirt}. The output SED is converged with respect to number of photon packages, maximum fraction of dust in one cell, minimum and maximum dust grid level.}
    \label{fig:skirt_convergence_tests}
\end{figure*}

We perform convergence tests on the parameters used in the radiative transfer. In Figure \ref{fig:skirt_convergence_tests}, we show the {\sc skirt-}predicted SED for halo A1 at $z=3$, for different {\sc skirt-} parameter choices. We re-run the radiative transfer using $5\times10^{6}$ photon packages (compared to the fiducial $1\times10^{6}$), maximum fraction of the dust mass per cell $5\times10^{-7}$ (compared to the fiducial $1\times10^{-6}$), minimum dust grid level 4 (compared to the fiducial 3), and maximum dust grid level 21 (compared to the fiducial 20). The output SED is converged with respect to all these parameters, at all wavelengths.

\bsp	
\label{lastpage}
\end{document}